\RequirePackage{rotating}
\documentclass[twocolumn, tighten]{aastex631}
\usepackage{amsmath}
\usepackage[caption=false]{subfig}
\usepackage{xfrac}
\usepackage{CJK}

\usepackage[T1]{fontenc}

\shortauthors{Huang et al.}

\begin{document}
\begin{CJK*}{UTF8}{gbsn} 
\title{High Resolution ALMA Observations of Richly Structured Protoplanetary Disks in $\sigma$ Orionis}

\correspondingauthor{Jane Huang}
\email{jane.huang@columbia.edu}
\author[0000-0001-6947-6072]{Jane Huang}
\affiliation{Department of Astronomy, Columbia University, 538 W. 120th Street, Pupin Hall, New York, NY 10027, USA}

\author[0000-0003-4142-9842]{Megan Ansdell}
\affiliation{NASA Headquarters, 300 E Street SW, Washington, DC 20546, USA}

\author[0000-0002-1899-8783]{Tilman Birnstiel}
\affiliation{University Observatory, Faculty of Physics, Ludwig-Maximilians-Universit\"at M\"unchen, Scheinerstr. 1, 81679 Munich, Germany}
\affiliation{Exzellenzcluster ORIGINS, Boltzmannstr. 2, D-85748 Garching, Germany}

 \author[0000-0002-1483-8811]{Ian Czekala}
\affiliation{School of Physics \& Astronomy, University of St. Andrews, North Haugh, St. Andrews KY16 9SS, UK}

\author[0000-0002-7607-719X]{Feng Long}
\altaffiliation{NASA Hubble Fellowship Program Sagan Fellow}
\affiliation{Lunar and Planetary Laboratory, University of Arizona, Tucson, AZ 85721, USA}

\author[0000-0001-5058-695X]{Jonathan Williams}\affiliation{Institute for Astronomy, University of Hawai‘i at M\=anoa, 2680 Woodlawn Dr., Honolulu, HI 96822, USA}

\author[0000-0002-8537-9114]{Shangjia Zhang}
\altaffiliation{NASA Hubble Fellowship Program Sagan Fellow}
\affiliation{Department of Astronomy, Columbia University, 538 W. 120th Street, Pupin Hall, New York, NY 10027, USA}
\affiliation{Department of Physics and Astronomy, University of Nevada, Las Vegas, 4505 S. Maryland Pkwy, Las Vegas, NV, 89154, USA}
\affiliation{Nevada Center for Astrophysics, University of Nevada, Las Vegas, Las Vegas, NV 89154, USA}

\author[0000-0003-3616-6822]{Zhaohuan Zhu}
\affiliation{Department of Physics and Astronomy, University of Nevada, Las Vegas, 4505 S. Maryland Pkwy, Las Vegas, NV, 89154, USA}
\affiliation{Nevada Center for Astrophysics, University of Nevada, Las Vegas, Las Vegas, NV 89154, USA}

\begin{abstract}
ALMA has detected substructures in numerous protoplanetary disks at radii from a few to over a hundred au. These substructures are commonly thought to be associated with planet formation, either by serving as sites fostering planetesimal formation or arising as a consequence of planet-disk interactions. Our current understanding of substructures, though, is primarily based on observations of nearby star-forming regions with mild UV environments, whereas stars are typically born in much harsher UV environments, which may inhibit planet formation in the outer disk through external photoevaporation.  We present high resolution ($\sim8$ au) ALMA 1.3 mm continuum images of eight disks in $\sigma$ Orionis, a cluster irradiated by an O9.5 star. Gaps and rings are resolved in the images of five disks. The most striking of these is SO 1274, which features five gaps that appear to be arranged nearly in a resonant chain. In addition, we infer the presence of gap or shoulder-like structures in the other three disks through visibility modeling. These observations indicate that substructures robustly form and survive at semi-major axes of several tens of au or less in disks exposed to intermediate levels of external UV radiation as well as in compact disks. However, our observations also suggest that disks in $\sigma$ Orionis are mostly small and thus millimeter continuum gaps beyond a disk radius of 50 au are rare in this region, possibly due to either external photoevaporation or age effects. 
\end{abstract}


\section{Introduction} \label{sec:intro}
The high detection rates of substructures such as gaps and rings in high-resolution ALMA millimeter continuum observations of Class II protoplanetary disks have profoundly altered our understanding of planet formation \citep[e.g.,][]{2015ApJ...808L...1A, 2018ApJ...869L..41A, 2018ApJ...869L..42H, 2018ApJ...869...17L, 2021MNRAS.501.2934C, 2024ApJ...966...59S}. Models have long predicted that massive protoplanets can open gaps in disks \citep[e.g.,][]{1986ApJ...307..395L, 1999ApJ...514..344B, 2000MNRAS.318...18N}, but prior to the advent of ALMA, most disks were assumed to have smooth surface density profiles, with a small fraction of older disks featuring inner cavities \citep[e.g.,][]{1998ApJ...500..411D, 2009ApJ...700.1502A}. ALMA's millimeter continuum observations are used to probe thermal emission from roughly millimeter-sized dust grains in disk midplanes. 
The substructures observed by ALMA at a range of disk radii and ages are commonly hypothesized to be the result of planet-disk interactions, implying that giant planet formation begins early and can occur readily even at semi-major axes of tens to more than a hundred au \citep[e.g.,][]{2015ApJ...806L..15K, 2018ApJ...869L..47Z, 2018ApJ...866..110D}. This interpretation has been bolstered by direct imaging detections of several giant protoplanets and protoplanet candidates and detections of non-Keplerian gas motion inside disk gaps \citep[e.g.,][]{2018AA...617A..44K, 2019NatAs...3.1109P, 2019Natur.574..378T, 2022NatAs...6..751C, 2023MNRAS.522L..51H}. Alternatively, some works have explored whether instead of being the outcome of planet formation, disk rings are sites that concentrate solids sufficiently to trigger planetesimal formation \citep[e.g.,][]{2020AA...638A...1M, 2022NatAs...6..357I, 2022ApJ...937...95L}. In this case, these substructures may arise due to processes such as dust accumulation at snowlines \citep[e.g.,][]{2015ApJ...806L...7Z, 2016ApJ...821...82O, 2017ApJ...845...68P}, zonal flows \citep[e.g.,][]{2009ApJ...697.1269J, 2014ApJ...796...31B}, surface density enhancements at the edges of dead zones \citep[e.g.,][]{2015AA...574A..10L,2015AA...574A..68F}, disk winds \citep[e.g.,][]{2018MNRAS.477.1239S}, or infall-driven instabilities \citep[e.g.,][]{2015ApJ...805...15B, 2022ApJ...928...92K}. Irrespective of their origins, substructures are thought to play an essential role in disk evolution through their influence on the radial transport of material through the disk, chemistry, and temperature \citep[e.g.,][]{2012AA...538A.114P, 2020ApJ...905...68A,2023ApJ...957L..22B}. 

However, one of the significant biases of published high-resolution millimeter wavelength disk observations is that they have principally targeted nearby ($d<200$ pc) star-forming regions, such as Taurus, Ophiuchus, Lupus, Chamaeleon, and Upper Sco \citep[see, e.g.,][and references therein]{2023ASPC..534..423B}. The disks targeted in these nearby regions are not necessarily representative of typical planet formation environments. Stars are often born in close proximity to O stars and thus exposed to external far ultraviolet (FUV) radiation fields of order $10^3-10^4$ $G_0$ \citep{2008ApJ...675.1361F, 2020MNRAS.491..903W}, where $G_0=1.6\times10^{-3}$ erg s$^{-1}$ cm$^{-2}$ is the \citet{1968BAN....19..421H} field. Consequently, the Solar System is thought to have been likely to form in such an environment \citep{2010ARAA..48...47A}. In contrast, estimates for the external FUV radiation fields of disks in nearby star-forming regions are generally on the order of $10^0-10^2$ $G_0$ \citep[e.g.,][]{2010ApJ...715.1370G, 2016ApJ...832..110C, 2020AA...640A...5T}. Upper Sco is an OB region that has undergone expansion \citep{2021MNRAS.507.1381S}, so it is possible that its disks were exposed to higher external UV radiation in the past. However, since high resolution ALMA studies of Upper Sco have focused on the most massive disks \citep{2018ApJ...869L..41A, 2023AA...670L...1S}, they have likely still been biased against disks that have experienced strong external UV radiation in the past.   

Models have demonstrated that external UV radiation can exert a significant influence on disk structure and therefore shape the properties of resulting planetary systems. \citet{2022MNRAS.515.4287W} found that external FUV radiation fields as low as $\sim100$ $G_0$  can have a significant effect on planetary growth and migration. Strong external UV radiation can drive disk mass loss through external photoevaporation, leading to smaller disks and shorter lifetimes \citep[e.g.,][]{1998ApJ...499..758J, 1999ApJ...515..669S}. Consequently, the occurrence rate of giant planets is expected to be lower around stars exposed to stronger external UV fields \citep[e.g.,][]{2000AA...362..968A, 2022MNRAS.515.4287W}. The migration behavior of protoplanets is also sensitive to external photoevaporation \citep[e.g.,][]{2004MNRAS.347..613V, 2022MNRAS.515.4287W}. 

The star-forming regions that feature a large number of stars currently exposed to high ($\gtrapprox 10^3$  $G_0$) external UV radiation are located at distances of 400 pc or beyond \citep[see, e.g., review by][]{2022EPJP..137.1132W}. As the nearest one of these, the Orion Molecular Cloud Complex ($d\sim400$ pc) has frequently been targeted for studies of the influence of external photoevaporation on disk populations. Hubble Space Telescope optical images of proplyds in the Trapezium showed ionization fronts due to radiation from the nearby O star $\theta^1$ Ori C \citep{1993ApJ...410..696O, 1994ApJ...436..194O}. ALMA observations indicate that the disk size distributions in the Orion Nebula Cluster (ONC) and the OMC1 cloud are shifted toward smaller radii compared to nearby star-forming regions, with some combination of external photoevaporation and dynamical truncation possibly setting disk sizes \citep{2018ApJ...860...77E, 2021ApJ...923..221O}. Millimeter wavelength surveys have also shown that disk masses tend to decrease as local FUV field strengths increase in the ONC, $\sigma$ Orionis, L1641, and L1647, consistent with the behavior expected from external photoevaporation \citep{2014ApJ...784...82M, 2017AJ....153..240A, 2023AA...673L...2V}. 

The diminished masses and sizes of disks close to massive stars in Orion raise the question of whether they still commonly harbor the kinds of millimeter continuum disk substructures that are widespread in nearby star-forming regions. Substructures have been detected in a couple of individual high-resolution ALMA studies of massive disks in Orion, including V1247 Ori and GW Ori \citep{2017ApJ...848L..11K, 2020ApJ...895L..18B, 2020Sci...369.1233K}. Most ALMA surveys targeting Orion, though, have had spatial resolutions coarser than 20 au \citep[e.g.,][]{2017AJ....153..240A, 2018ApJ...860...77E, 2020AJ....160..248A, 2020AA...640A..27V, 2023AA...673L...2V, 2023ApJ...954..127B}, which is wider than the typical scales of disk substructures resolved in nearby star-forming regions \citep[e.g.,][]{2018ApJ...869...17L, 2018ApJ...869L..42H, 2021MNRAS.501.2934C}. The highest resolution ALMA survey of disks in Orion published thus far has been a $0\farcs03$ ($\sim12$ au) survey of the ONC and OMC1 by \citet{2021ApJ...923..221O}. Some of their disk images hint at the presence of substructures, but the bright large-scale emission in these regions poses a challenge to characterizing disk morphology. 

The $\sigma$ Orionis cluster presents a prime opportunity to investigate the properties of disks exposed to strong external UV radiation. The cluster is generally estimated to be about $3-5$ Myr old, albeit with large uncertainties \citep{2002AA...384..937Z, 2002AA...382L..22O, 2004MNRAS.347.1327O, 2019AA...629A.114C}. The cluster is strongly irradiated by an O9.5 star within the eponymous multiple star system $\sigma$ Orionis \citep{1953ApJ...117..313J, 1967PASP...79..433G, 1998ASPC..154.1793W}. (To avoid confusion, we will henceforth refer to the cluster as $\sigma$ Orionis and the star as $\sigma$ Ori). With $A_V$ values generally $<1$, the extinction toward this region is low compared to other parts of Orion \citep{1968ApJ...152..913L,2004AN....325..705B}, so contamination from large-scale emission does not present a problem for millimeter continuum disk imaging. To examine the structure of disks in this cluster in greater detail, we used ALMA to observe a sample of eight disks at a resolution of $0\farcs02$ ($\sim8$ au), improving upon previous observations by an order of magnitude. Section \ref{sec:sample} provides an overview of the selected targets. The observations and data reduction are described in Section \ref{sec:observations}, while modeling and analysis are presented in Section \ref{sec:analysis}. The results are discussed in Section \ref{sec:discussion} and summarized in Section \ref{sec:summary}. 



\section{Sample overview\label{sec:sample}}
Our eight targets were selected from the \citet{2017AJ....153..240A} ALMA 1.3 mm continuum survey of Class II disks in $\sigma$ Orionis, which imaged disks at a resolution of $\sim0\farcs25$ ($\sim100$ au). Disk classifications were based on the Spitzer survey by \citet{2007ApJ...662.1067H}. While one of the targets, SO 1153, was categorized by \citet{2007ApJ...662.1067H} as Class I, \citet{2017AJ....153..240A} included it in their survey because its colors were borderline between Class I and Class II. \citet{2017AJ....153..240A} and \citet{2023AA...679A..82M} did not note any obvious envelope emission in their millimeter continuum and $^{12}$CO observations of SO 1153, although the disk is only moderately resolved. 

The sample was restricted to stars with estimated stellar masses between $\sim0.4-1$ $M_\odot$, corresponding to the mass range for which disks have been best characterized with high resolution ALMA observations in nearby star-forming regions \citep[e.g.,][]{2018ApJ...869L..41A, 2018ApJ...869...17L, 2021MNRAS.501.2934C}. Because disks in $\sigma$ Orionis had not previously been observed at high resolution, we set a conservative flux cutoff of 1.5 mJy to ensure a reasonable signal-to-noise ratio. This cutoff was determined by generating synthetic ALMA images of disks based on the best-fit models of the Taurus disks in \citet{2018ApJ...869...17L}, but at a distance of 400 pc, to test the detectability of analogous substructures in $\sigma$ Orionis. 

Given the flux and stellar mass constraints, we then chose our targets to span a range of projected separations from $\sigma$ Ori. There is some uncertainty in the literature regarding the distance to $\sigma$ Ori. The O9.5 star does not have a \textit{Gaia} parallax due to its extreme brightness. Using infrared interferometry, \citet{2016AJ....152..213S} measured a distance of 387.5$\pm1.3$ pc to $\sigma$ Ori. However, the median \textit{Gaia} distance to members of $\sigma$ Orionis is 402 pc \citep{2024AA...686A.161Z}. \citet{2024AA...686A.161Z} commented that there may be some systematic offset between distances measured from interferometry and from \textit{Gaia}, since $\sigma$ Ori is thought to be at the center of its cluster. Given this uncertainty, we refer to projected rather than absolute separations from $\sigma$ Ori in this work. The projected separations of the targets range from 0.68 to 2.73 pc, which would correspond to FUV fields of $\sim200-2600$ $G_0$ (see Table \ref{tab:hoststars} and \citealt{2023AA...679A..82M}). The FUV field estimates from \citet{2023AA...679A..82M} are based on projected separations, so they should be considered upper bounds. The host star properties of the selected targets are listed in Table \ref{tab:hoststars}.

\begin{deluxetable*}{lccccccc}
\tablecaption {Host star properties \label{tab:hoststars}}
\tablehead{\colhead{Name} &\colhead{2MASS designation}&\colhead{SpT} & \colhead{M$_\ast$} &\colhead{$L_\ast$}&\colhead{Distance}&\colhead{Projected separation from $\sigma$ Ori}&\colhead{FUV field}\\
&&&($M_\odot$)&($L_\odot$)&(pc)&(pc)&\colhead{($\log G_0$)}}
\startdata
SO 662 & J05384027-0230185 &K7 & 0.64 & 0.68&$394.9\pm3.1$ & 0.68 & 3.41\\
SO 844 & J05390136-0218274 & M1 & 0.44 &0.62&$408.4\substack{+3.4 \\ -3.6}$&2.11 & 2.42 \\
SO 897 & J05390760-0232391 & K6 & 0.7 & 0.85&$375.5\substack{+8.0 \\ -6.8}$ & 0.77 & 3.29\\
SO 984 & J05391883-0230531& K7 & 0.64&0.72&$403.5\substack{+3.0 \\- 3.4}$&1.16 & 2.93\\
SO 1036 & J05392519-0238220 &M0 & 0.59 &0.53&$388.7\substack{+2.9 \\- 3.4}$&1.21 & 2.91\\
SO 1152 & J05393938-0217045 &M0 & 0.58 &0.61& $391.3\substack{+4.2 \\- 3.3}$&2.73 & 2.20\\ 
SO 1153 & J05393982-0231218 & K5 & 0.9 & 0.33&$390.4\substack{+3.7 \\- 3.6}$&1.70 & 2.61\\
SO 1274 & J05395465-0246341&K7 & 0.64& 0.68&$400.1\substack{+3.7 \\- 2.9}$&2.39 & 2.31\\
\enddata 
\tablerefs{Spectral type, stellar mass, stellar luminosity, and external FUV radiation field values come from \citet{2023AA...679A..82M}. Our FUV field values are rescaled from \citet{2023AA...679A..82M} because they calculate projected separations using the distances to the individual sources, whereas we use the median distance to the cluster of 402 pc \citep{2024AA...686A.161Z} so that the projected distances scale linearly with the angular separations.} Stellar distances come from \citet{2021AJ....161..147B}, which is based on data from \citet{2021AA...649A...1G}.
\end{deluxetable*}

\begin{figure*}
\begin{center}
\includegraphics{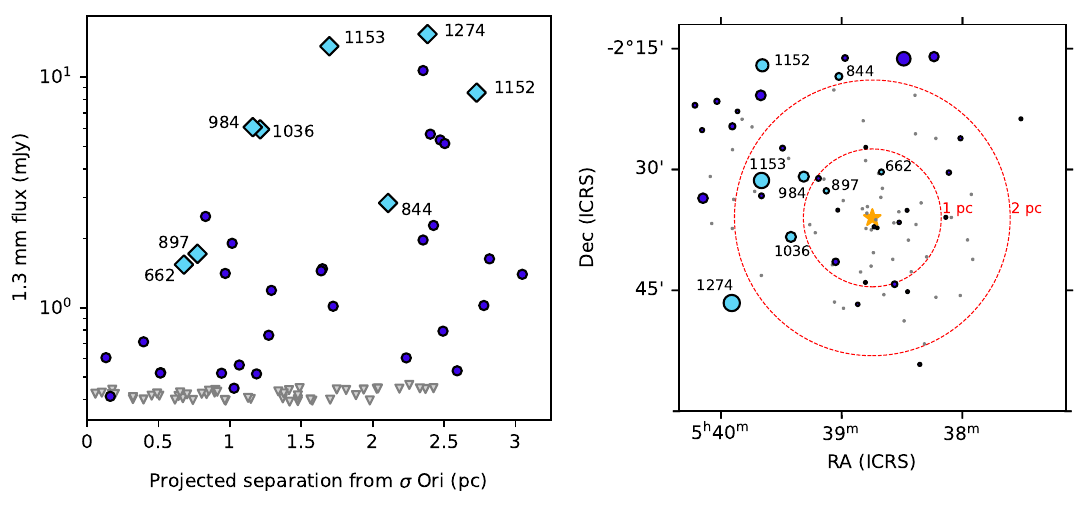}
\end{center}
\caption{Left: A comparison of the 1.3 mm fluxes and projected separations of the sources observed in this work (light blue diamonds) and other disk detections (purple dots) and upper limits (gray triangles) in $\sigma$ Orionis from \citet{2017AJ....153..240A}. The flux values used are all from \citet{2017AJ....153..240A} for consistency. Right: A plot of the coordinates of the $\sigma$ Orionis sources observed by  \citet{2017AJ....153..240A}. Light blue dots correspond to sources observed in this work, while purple dots correspond to other disk detections and gray dots correspond to non-detections from \citet{2017AJ....153..240A}. The position of $\sigma$ Ori is marked by an orange star. The sizes of the markers for the detected sources are scaled by the 1.3 mm flux. The dashed circles mark projected separations of 1 and 2 pc, respectively.\label{fig:Cycle9targets}}
\end{figure*}

 No obvious substructures are visible in the millimeter continuum images in \citet{2017AJ....153..240A} or \citet{2023AA...679A..82M}, which reobserved some of the targets from \citet{2017AJ....153..240A} at similar resolution but higher sensitivity. \citet{2024AA...685A..54V} used VLT/SPHERE to image the SO 1274 disk in infrared scattered light, which traces micron/sub-micron-sized dust grains in the upper layers of disks. In the SPHERE image, the disk appears faint, with no apparent substructure.  However, among our targets, \citet{2007ApJ...662.1067H} identified SO 897 as a transition disk candidate based on its infrared colors. 
 
\begin{figure}
\begin{center}
\includegraphics{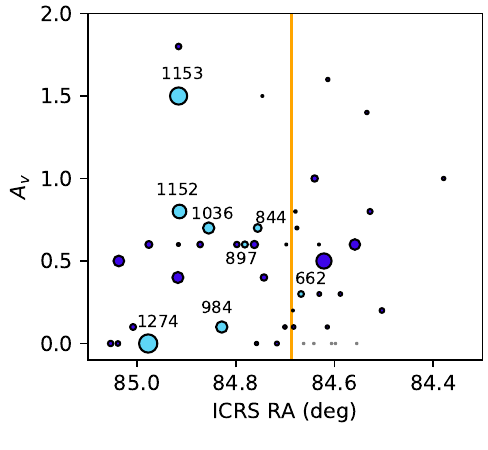}
\end{center}
\caption{$A_V$ of stars in $\sigma$ Orionis observed by \citet{2023AA...679A..82M}, plotted as a function of right ascension. The vertical orange line marks the R.A. of $\sigma$ Ori. Light blue dots correspond to sources observed in this work, while purple dots correspond to other disk detections and gray dots correspond to non-detections. The sizes of the markers for the detected sources are scaled by the 1.3 mm flux measured by \citet{2023AA...679A..82M}.\label{fig:extinction}}
\end{figure}

 The fluxes and projected separations of the targets are plotted relative to the rest of the disks detected in $\sigma$ Orionis in Figure \ref{fig:Cycle9targets}. While our flux cutoff biases our sample towards brighter sources, our target fluxes still span an order of magnitude. A two-dimensional plot of the positions of the targets with respect to $\sigma$ Ori is also shown in Figure \ref{fig:Cycle9targets}. The targets are primarily to the east of $\sigma$ Ori, which is a consequence of members of the cluster being preferentially located to the east of $\sigma$ Ori \citep{2008MNRAS.383..375C}. \citet{2008MNRAS.383..375C} speculated that the asymmetric distribution of stars in $\sigma$ Orionis resulted from variations in the dust surface density in the molecular cloud at the start of star formation. \citet{2023AA...679A..82M} measured the $A_V$ and 1.3 mm continuum disk fluxes for 50 stars in $\sigma$ Orionis; the $A_V$ values as a function of right ascension are plotted in Figure \ref{fig:extinction}. The $A_V$ values are $<1$ for most sources and exhibit a similar spread to the east and west of $\sigma$ Ori. However, given the intermediate age of the cluster, the present day spatial variation (or lack thereof) in $A_V$ may not necessarily reflect the relative extinction levels between the east and west sides of $\sigma$ Orionis early on in the cluster's history.

\section{Observations and Data Reduction}\label{sec:observations}
Long-baseline 1.3 mm continuum observations of the eight disks were obtained by ALMA program 2022.1.00728.S (PI: J. Huang). For all observations, the correlator was configured with four spectral windows (SPWs) centered at 224, 226, 240, and 242 GHz, respectively. Each SPW had a bandwidth of 2 GHz divided into 128 channels. The time on target for each execution block (EB) was 29 minutes 40 seconds. Three EBs were used to observe SO 1274, while other targets were observed for two EBs each. The quasars J0423-0120, J0532-0307, and J0529-0519 were used as calibrators for all observations. Table \ref{tab:observations} lists the observing date, baseline lengths, and number of antennas for each EB. Calibrated measurement sets were produced by ALMA staff with the standard ALMA pipeline in \texttt{CASA 6.4.1} \citep{2022PASP..134k4501C}.

\begin{deluxetable*}{lccc}
\tablecaption {Description of execution blocks in ALMA program 2022.1.00728.S  \label{tab:observations}}
\tablehead{\colhead{Source} &\colhead{Date}&\colhead{Baseline lengths} & \colhead{Number of antennas}}
\startdata
SO 662 & 2023 July 26 &256 m - 15.2 km  & 43 \\
 & 2023 July 27 &230 m - 16.2 km  & 45 \\
\hline
SO 844 & 2023 July 31 & 230 m - 16.2 km & 43 \\
& 2023 August 02 & 230 m - 16.2 km & 46 \\
\hline
SO 897 & 2023 July 28 & 230 m - 16.2 km & 48\\
& 2023 July 31 & 230 m - 16.2 km & 46 \\
\hline
SO 984 & 2023 August 06 & 230 m - 16.2 km & 44  \\
 & 2023 August 07 & 230 m - 16.2 km& 48  \\
\hline
SO 1036 & 2023 August 04 &230 m - 16.2 km & 43  \\
&2023 August 05 &230 m - 16.2 km & 44 \\
\hline
SO 1152 & 2023 August 02 &230 m - 16.2 km&46  \\ 
 & 2023 August 03 &230 m - 16.2 km&46  \\ 
\hline
SO 1153 & 2023 August 01 & 230 m - 16.2 km & 43 \\
&2023 August 04 & 230 m - 16.2 km & 43 \\
\hline
SO 1274 & 2023 July 19 &230 m - 15.2 km&46\\
 & 2023 July 27 &230 m - 16.2 km&45\\
 & 2023 July 29 &230 m - 16.2 km&44 \\
\enddata 
\end{deluxetable*}

To provide $uv$ coverage at shorter distances, we retrieved lower-resolution 1.3 mm observations of our targets from program 2016.1.00447.S (PI: J. Williams) through the ALMA archive. These observations were first published in \citet{2023AA...679A..82M}, which describes them in more detail. In brief, each target was observed with baselines ranging from $\sim15$ m to 2.6 km over the course of eight EBs, with a total time on-target of $\sim9$ minutes. The raw data were calibrated with the \texttt{CASA} 4.7.2 pipeline. Three of the SPWs covered $^{12}$CO, $^{13}$CO, and C$^{18}$O $J=2-1$; the $^{12}$CO images are presented in \citet{2023AA...679A..82M}. 

Subsequent processing of the new and archival data was performed with \texttt{CASA} 6.5. First, we flagged channels where CO line emission might be present in the archival short-baseline data and averaged the channels to create pseudo-continuum visibilities for each target. No self-calibration was applied to SO 897 (the faintest disk in the sample) due to its low signal-to-noise ratio. For the other disks, phase self-calibration was first performed separately on each EB from the archival short-baseline dataset with a solution interval spanning all scans. Images of each EB were produced with the H\"ogbom CLEAN algorithm \citep{1974AAS...15..417H} as implemented in the \texttt{tclean} task. (Since the disks are either marginally resolved or unresolved in the short-baseline observations, multi-scale CLEAN is not necessary). As in \citet{2018ApJ...869L..41A}, the disk centers were determined by fitting each image with a two-dimensional Gaussian using the \texttt{imfit} task. The EBs were then aligned with one another using the \texttt{phaseshift} and \texttt{fixplanets} tasks such that the disk emission was centered at the phase center. For each disk except SO 897 and SO 844 (which had insufficient S/N), the EBs were then imaged together and phase self-calibration was performed with scan-length solution intervals. Finally, amplitude self-calibration was performed with scan-length solution intervals on the combined EBs for all disks except SO 662, SO 844, and SO 897, which are the three faintest disks in the sample. 

We then created channel-averaged measurement sets from each EB of the long-baseline observations and imaged them separately using multi-scale CLEAN \citep{2008ISTSP...2..793C}. Phase self-calibration was performed on the individual EBs for SO 1153, the brightest disk in the sample. Self-calibration was tested on the other sources, but did not improve the images. For SO 1274, SO 1153, and SO 1036, the execution blocks were aligned with the same procedure as the short-baseline observations. The disk emission appeared to be well-centered for the other sources, so no phase shift was applied, but the phase centers were relabeled with \texttt{fixplanets} to match the short-baseline observations. The short-baseline and long-baseline observations were then combined and imaged together with multi-scale CLEAN and a robust value of 0.5. For SO 1153, phase self-calibration was performed on the combined observations with a scan-length solution interval. Finally, all of the disk images were primary-beam-corrected. The resulting image properties are listed in Table \ref{tab:imageproperties}. The continuum visibilities and images can be downloaded from \url{https://zenodo.org/records/13821034}.

\begin{deluxetable}{lccc}
\tablecaption {Image properties \label{tab:imageproperties}}
\tablehead{\colhead{Source} &\colhead{Synthesized beam}&\colhead{rms} & \colhead{Peak Intensity}\\
\colhead{} &\colhead{(mas $\times$ mas ($^\circ$))}&\colhead{($\mu$Jy beam$^{-1}$)} & \colhead{(mJy beam$^{-1}$)}}
\startdata
SO 662 &$22\times 20$ ($72.8$)&10&0.42 \\
SO 844 & $22\times 19$ ($-77.1$)&10&0.20\\
SO 897 & $24\times 20$ ($77.4$)& 10 & 0.33\\
SO 984 &$22\times 20$ ($-85.0$) &11&0.38\\
SO 1036 & $24\times 19$ ($-60.8$)&12 &0.46\\
SO 1152 & $23\times 21$ ($-82.1$) & 9 & 0.27\\ 
SO 1153 &$23\times 20$ ($-70.5$) &11& 0.84 \\
SO 1274 & $23\times 21$ ($87.0$)& 9&0.41 \\
\enddata 
\end{deluxetable}

\begin{figure*}
\begin{center}
\includegraphics{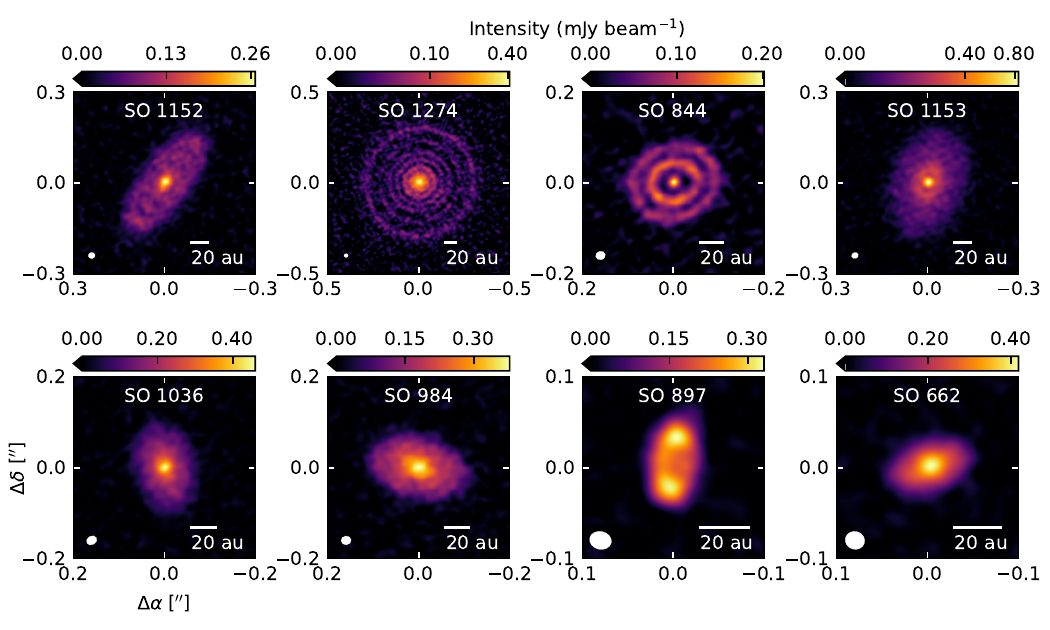}
\end{center}
\caption{1.3 mm ALMA continuum images of the disks, sorted in descending order of projected separation from $\sigma$ Ori. The synthesized beam is shown in the lower left corner of each panel. An asinh stretch is used for some of the disks in order to show the faint emission at larger radii more clearly. \label{fig:continuumgallery}}
\end{figure*}

\begin{figure*}
\begin{center}
\includegraphics{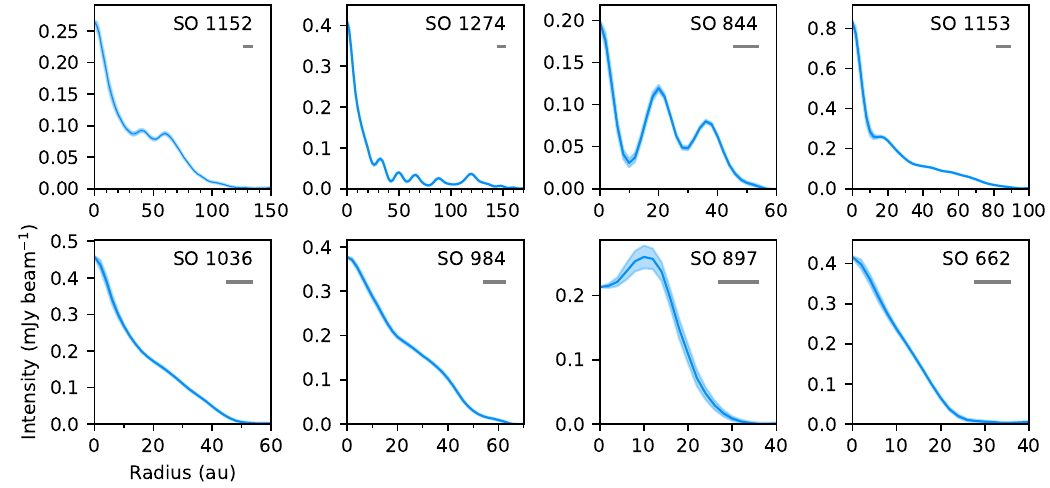}
\end{center}
\caption{Azimuthally averaged, deprojected radial profiles of the disks produced from the CLEAN images. The blue shaded ribbon shows the $1\sigma$ error, which is estimated by taking the standard deviation of the pixels in each radial bin and dividing by the square root of the number of beams spanned by the corresponding annulus. The gray bar in the upper right corner of each panel denotes the major axis of the synthesized beam.  \label{fig:radialprofiles}}
\end{figure*}

\section{Analysis\label{sec:analysis}}
\subsection{Overview of disk emission morphology}
Continuum images are shown in Figure \ref{fig:continuumgallery}, while azimuthally averaged, deprojected radial profiles are shown in Figure \ref{fig:radialprofiles}. Annular gaps and rings are visible in the SO 1152, SO 1274, SO 844, and SO 1153 disks, while a small inner cavity is detected in the SO 897 disk. Some of the substructures appear to be slightly hexagonal due to the shape of the point spread function \citep[see, e.g.,][]{2018ApJ...869L..41A}. No gaps are immediately evident in the SO 1036, SO 984, and SO 662 disks, but the radial profiles of SO 1036 and SO 984 exhibit subtle slope changes. 

\begin{figure}
\begin{center}
\includegraphics{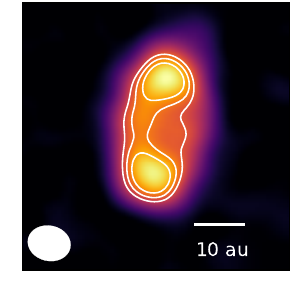}
\end{center}
\caption{A zoomed-in view of SO 897, showing the emission asymmetry across the major axis. Contours are drawn at 0.6, 0.7, and 0.8$\times$ the peak intensity. \label{fig:SO897}}
\end{figure}

The disks generally appear to be axisymmetric. However, SO 897 is slightly brighter (by $\sim15\%$) on its eastern side compared to its western side (Figure \ref{fig:SO897}). Given the high inclination of the disk, this brightness asymmetry is likely a consequence of the viewing geometry. For disks at higher inclination, the far side can appear brighter either due to a puffed-up cavity wall \citep[e.g.,][]{2010ARAA..48..205D, 2024MNRAS.532.1752R} or a geometrically thick disk \citep[e.g.,][]{2023ApJ...951....8O}. 

\subsection{Parametric intensity profile modeling}

Modeling disk emission in the $uv$ plane is often an effective technique for inferring the presence of substructures that are not readily visible in CLEAN images \citep[e.g.,][]{2016ApJ...818L..16Z, 2020MNRAS.495.3209J, 2023AJ....166..184M}. We thus fit the observations with parametric intensity models in order to infer the radial intensity profiles of our targets. For all disks except SO 897, we assumed that their radial intensity profiles can be described as the sums of a Gaussian component centered at $r=0$ and $N$ additional Gaussian components with offsets in the radial directions (these are commonly referred to as Gaussian rings, but $N$ does not necessarily correspond to the number of rings visible in the total intensity profile if two or more of the Gaussian components are closely overlapping):
\begin{equation}\label{eq:radprofile}
I(r) = A_0\exp\left(-\frac{r^2}{2\sigma_0^2}\right)+ \sum_{i=1}^{N} A_i \exp\left(-\frac{(r-r_i)^2}{2\sigma_i^2}\right),
\end{equation}
where $r$ is the disk radius in au in cylindrical coordinates. Expressions of this form have been shown to reproduce high-resolution ALMA observations of disks well \citep[e.g.,][]{2018ApJ...869L..48G, 2018ApJ...869L..49I}. The value of $N$ for each disk was initially chosen based on the number of rings or extended emission tails visible in the radial profiles extracted from ALMA images, and higher values of $N$ were tested to determine whether they better reproduce the disk emission. Since adding more parameters can result in overfitting the data, we calculated the Bayesian information criterion \citep{Schwarz1978} to determine which model to select (Appendix \ref{sec:modelselection}). 

Given a model radial intensity profile, we then used the \texttt{mpol} package \citep{2023PASP..135f4503Z, mpol} to generate a model disk image with some position angle (P. A.), inclination (\textit{i}), east-west offset from the phase center ($\Delta x$), and north-south offset from the phase center ($\Delta y$), under the assumption that the disk is geometrically thin. Thus, $3N+6$ free parameters are required to specify a disk model fully.  

For SO 897, which has an inner cavity, we adopted the following radial intensity profile: 
\begin{equation}
I(r) = A_1 \exp\left(-\frac{(r-r_1)^2}{2\sigma_1^2}\right). 
\end{equation}

Given the observed emission asymmetry (Figure \ref{fig:SO897}), we did not assume that the emission comes from a flat surface ($z=0$ in disk coordinates). Instead, we assumed that the emission comes from a flared surface of the form $z(\rho) = z_0 \left(\frac{\rho}{1''}\right)^\phi$, where $\rho$ is the disk radius in arcseconds and  $z_0$ and $\phi$ are free parameters. We then used the \texttt{eddy} package \citep{2019JOSS....4.1220T} to transform disk coordinates to sky coordinates and \texttt{mpol} to generate a model image for some P.A., $i$, $\Delta x$, and $\Delta y$. Thus, 9 free parameters total are required to specify the model for SO 897. 

For all disks, we then used \texttt{mpol} to generate model visibilities $\mathcal{V}_\text{mod}$ from the model images at the same $uv$ coordinates as the observed visibilities $\mathcal{V}_\text{obs}$. The log likelihood (up to a constant) is
\begin{equation}
\ln \mathcal{L} = -\frac{1}{2} \sum_{i=1}^{n} w_i \left| \mathcal{V}_\text{obs,i}-\mathcal{V}_\text{mod,i}\right|^2,
\end{equation}
where $w$ denotes the visibility weights and $n$ is the total number of visibilities. In general, the absolute scaling of weights in calibrated measurement sets delivered by ALMA may not be correct (see, for example, \url{https://casaguides.nrao.edu/index.php/DataWeightsAndCombination} and \citet{2013ApJ...767..132H}). Hence, we used the procedure described in \citet{2023PASP..135f4503Z} and implemented in the \texttt{visread} package \citep{ian_czekala_2021_4432520} to correct the scaling. For each disk, the CLEAN model was subtracted from the visibilities, and then a Gaussian was fit to the scatter in the residual visibilities normalized by $\sigma_V = \sqrt{\frac{1}{w}}$. The weights were then rescaled by a factor of $\frac{1}{f^2}$, where $f$ was the standard deviation of the best-fit Gaussian. The value of $f$ was typically $\sim2$. 

Gaussian priors were specified for  $\log A_0$, $\log \sigma_0$, $\log A_i$, $\log \sigma_i$, $r_i$, $i$, $z_0$, $\phi$, P.A., $\Delta x$, $\Delta y$. Their means and standard deviations were set based on visual inspection of the observed images and radial profiles. 

The posterior distributions were estimated with the \texttt{pyro} \citep{JMLR:v20:18-403} implementation of the stochastic variational inference (SVI) algorithm \citep{hoffman2013}. By using parametric expressions to approximate the posterior distributions, SVI can be used to estimate the posterior much faster than Markov Chain Monte Carlo (MCMC) methods typically can, and is therefore particularly advantageous for high-dimensional models, where MCMC often struggles.\footnote{A demonstration of the application of SVI to parametric visibility modeling of the AS 209 disk and comparison to MCMC results can be found at \url{https://github.com/MPoL-dev/examples/blob/main/AS209-pyro-inference/pyro.ipynb}.} However, it should be kept in mind that MCMC methods are theoretically guaranteed to converge to the true posterior for sufficiently long runs, whereas SVI is an approximate method. We assumed that the posterior can be approximated as a multivariate normal distribution (a ``guide'' in the parlance of \texttt{pyro}). This assumption appears reasonable based on the posterior distributions derived from MCMC modeling of the radial profiles of circumstellar disks \citep[e.g.,][]{2015ApJ...801...59M, 2019ApJ...881...84S}. More generally, it is oftentimes the case that posterior distributions are approximately normal. Stochastic gradient descent was used to optimize the evidence lower bound (ELBO) in order to estimate the parameters of the guide. Using the Adam optimizer \citep{kingma2017} with a learning rate of 0.02, we found that 15,000 iterations were sufficient for the ELBO values to converge (note that these iterations are steps in the optimization routine and should not be confused with samples of the posterior).

\begin{figure*}
\begin{center}
\includegraphics{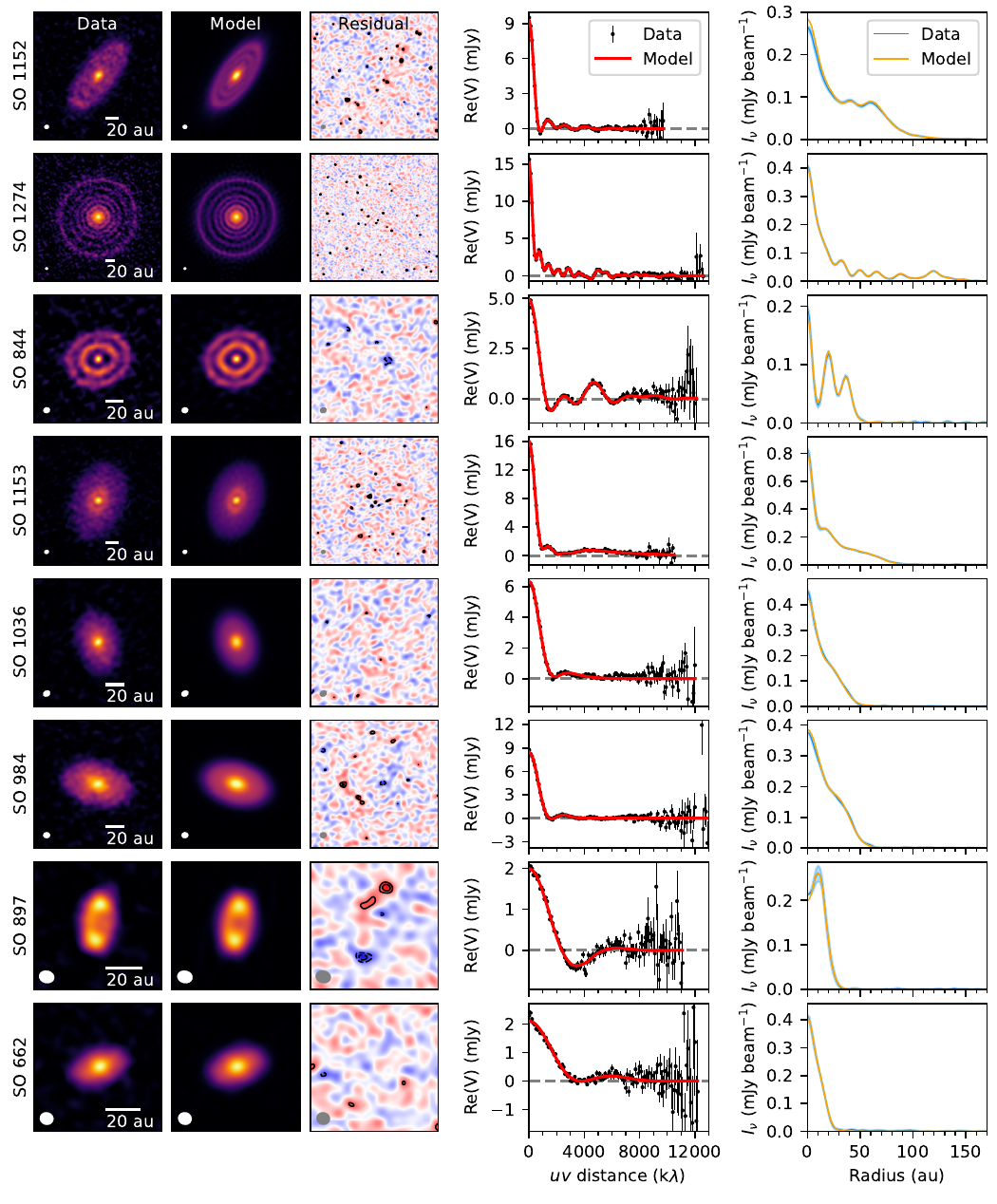}
\end{center}
\caption{A comparison of the parametric intensity models to the observations. Column 1: Observations (on the same color scale as Figure \ref{fig:continuumgallery}). Column 2: CLEAN images generated from the model visibilities, on the same color scale as column 1. Column 3: Residual images, with the color scale ranging from -5$\sigma$ (blue) to 5$\sigma$ (red), where $\sigma$ is the rms listed in Table \ref{tab:imageproperties}. Dashed contours correspond to the $[-4,-3]\sigma$ level and solid contours to the $[3,4]\sigma$ level. Column 4: A comparison of the deprojected, binned observed visibilities to the models. Column 5: A comparison of the deprojected, azimuthally averaged profiles made from the observed and model CLEAN images. \label{fig:modelcomparisons}}
\end{figure*}

For each disk, we then generated 2000 samples from the estimated posteriors and calculated the median value of each parameter and the $68\%$ confidence intervals. The model values are given in Appendix \ref{sec:modelparameters}. To check that the parametric intensity models reasonably describe the observations, we generated model visibilities from the median parameter values and then imaged them in the same manner as the observations. A comparison between the models and observations is shown in Figure \ref{fig:modelcomparisons}. The models reproduce the data well, with maximum residual levels at $\sim4\sigma$. 

\subsection{Substructure properties}
\subsubsection{Locations, widths, and depths}
\begin{figure*}
\begin{center}
\includegraphics{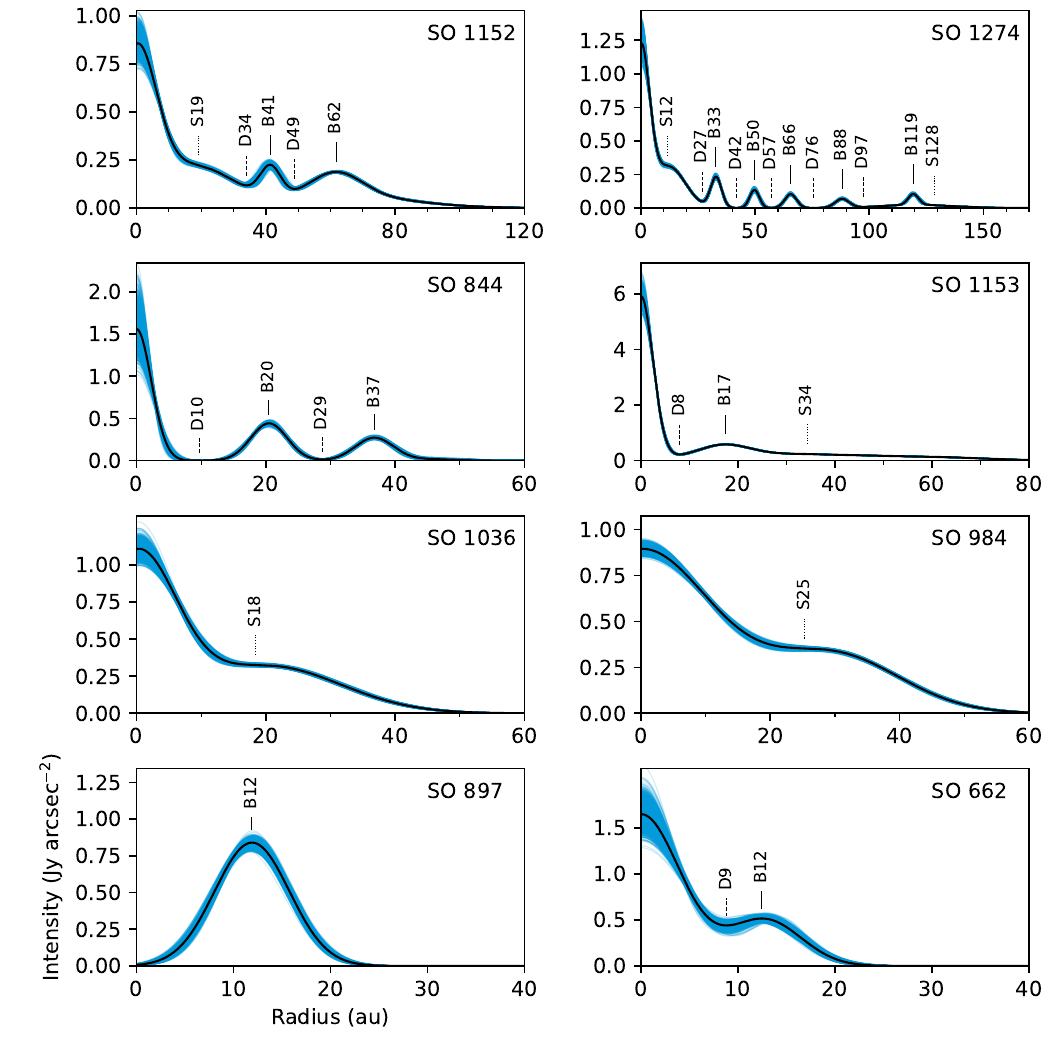}
\end{center}
\caption{Radial intensity profiles derived from parametric modelling and labelled with the locations of the substructures. The black curves correspond to profiles generated with the median values of the marginalized posterior distributions, while the blue curves show the results of 2000 draws from the posterior. \label{fig:annotatedprofiles}}
\end{figure*}
With the posterior samples from above, we then generated 2000 model radial intensity profiles for each disk (Figure \ref{fig:annotatedprofiles}). Substructures were identified in the following manner: A ring is defined to be a local maximum in the radial intensity profile (other than at the disk center), with its location corresponding to the radius at which the maximum occurs. Likewise, a gap is a local minimum in the profile, with the location determined by the radius where the minimum occurs. We follow the nomenclature of \citet{2018ApJ...869L..42H}, such that a ring is labelled by the letter ``B'' followed by the radial location rounded to the nearest au, and a gap is labelled by the letter ``D'' followed by the radial location rounded to the nearest au. A cavity is identified if the intensity at $r=0$ is less than the peak intensity. However, given that inner disks have now been detected in a number of cavities imaged at high resolution \citep[e.g.,][]{2020ApJ...892..111F}, the distinction between a cavity and a gap in some cases may be a matter of resolution. A shoulder is identified when a radial profile features three consecutive inflection points without a local maximum or minimum occurring between the first and last inflection point. In other words, the radial profile changes from being concave down to concave up and then concave down again without a ring or gap being present. The radial location is defined to be that of the middle inflection point. This definition is diagrammed in Appendix \ref{sec:shoulder}. The features that we label shoulders have sometimes been classified as gaps and rings in the literature \citep[e.g.,][]{2018ApJ...869L..42H} because unresolved gaps and rings can create shoulder-like features in radial profiles \citep[e.g.,][]{2016ApJ...818L..16Z}. However, for this work we elect to be more conservative in our definitions. The measured locations, widths, and depths of the substructures are provided in Table \ref{tab:substructureproperties}.

\begin{deluxetable}{lcccc}
\tablecaption {Substructure properties \label{tab:substructureproperties}}
\tablehead{\colhead{Source} &\colhead{Feature} &\colhead{Radial location}&\colhead{Width} & \colhead{Depth}\\
&&(au)&(au)}
\startdata
SO 662 &D9&$8.9\pm0.3$&$3.1\substack{+0.4\\ -0.5}$&$0.85\pm0.07$\\
&B12&$12.46\substack{+0.45\\-0.49}$&$3.5\substack{+0.4 \\ -0.6}$&-\\
\hline
SO 844 & D10&$9.7\pm0.4$&$12.6\pm0.3$&$0.0011\substack{+0.0008 \\ -0.0005}$\\
&B20&$20.48\substack{+0.11 \\ -0.12}$&$6.6\pm0.3$&-\\
&D29&$28.7\pm0.3$&$8.7\pm0.2$&$0.049\substack{+0.010 \\ -0.009}$\\
&B37&$36.8\pm0.2$&$6.8\pm0.3$&-\\
\hline
SO 897 & Cavity & - & $7.4\pm0.2$&$0.008\pm0.003$ \\
& B12&$11.9\pm0.1$&$9.0\pm0.3$&-\\
\hline
SO 984 &S25 &$25.3\pm0.2$&-&-\\
\hline
SO 1036 &S18 &$18.4\substack{+0.4 \\ -0.5}$&-&- \\
\hline
SO 1152 & S19 &$19.2\pm0.7$ &- & -\\
& D34 & $34.0\pm0.6$ & $11.2\pm0.6$ & $0.52\pm0.03$\\
&B41&$41.3\pm0.5$&$6.3\pm0.3$&-\\
&D49&$49.0\pm0.5$&$9.3\pm0.3$&$0.53\pm0.02$\\
&B62&$61.8\pm0.4$&$14.2\pm0.4$& - \\
\hline
SO 1153 & D8&$8.01\substack{+0.15 \\ -0.14}$&$6.10\substack{+0.13 \\ -0.14}$&$0.38\pm0.02$\\ 
&B17&$17.48\substack{+0.16\\-0.18}$&$11.3\pm0.2$&-\\
&S34&$34.3\pm0.4$&-&-\\
\hline
SO 1274 & S12 & $11.6\pm0.3$ &-&-\\
& D27 & $26.9\pm0.2$&$9.1\pm0.2$&$0.21\pm0.02$\\
&B33&$32.8\pm0.2$& $5.0\pm0.2$&-\\
&D42&$42.1\pm0.3$&$10.6\substack{+0.3\\-0.4}$&$0.0030\substack{+0.0018 \\ -0.0011}$\\
&B50&$49.7\pm0.2$&$4.9\pm0.3$&-\\
&D57&$57.0\pm0.4$&$9.9\pm0.4$&$0.007\substack{+0.004 \\ -0.003}$\\
&B66&$65.6\pm0.2$&$6.0\pm0.4$&-\\
&D76&$75.6\substack{+0.5 \\ -0.4}$&$15.2\substack{+0.4 \\ -0.5}$&$0.0046\substack{+0.0018 \\ -0.0014}$\\
&B88&$88.2\pm0.3$&$7.4\pm0.6$&-\\
&D97&$97.3\substack{+0.6\\-0.7}$&$26.2\pm0.6$&$0.070\substack{+0.008 \\ -0.007}$\\
&B119&$119.3\pm0.3$&$6.0\pm0.5$&-\\
&S128&$128.47\substack{+0.84\\-0.79}$&-&-
\enddata 
\end{deluxetable}

In addition to the substructures visible in the radial profiles of the CLEAN images (Figure \ref{fig:radialprofiles}), we infer the presence of a gap in the SO 662 disk and shoulders in the SO 984, SO 1036, SO 1152, SO 1153, and SO 1274 disks. To check that the inferred substructures are not merely an artifact from our choice of parametrization, we also modelled the visibilities with the non-parametric modeling code \texttt{frank} \citep{2020MNRAS.495.3209J} (Appendix \ref{sec:frank}). 

\begin{figure*}
\begin{center}
\includegraphics{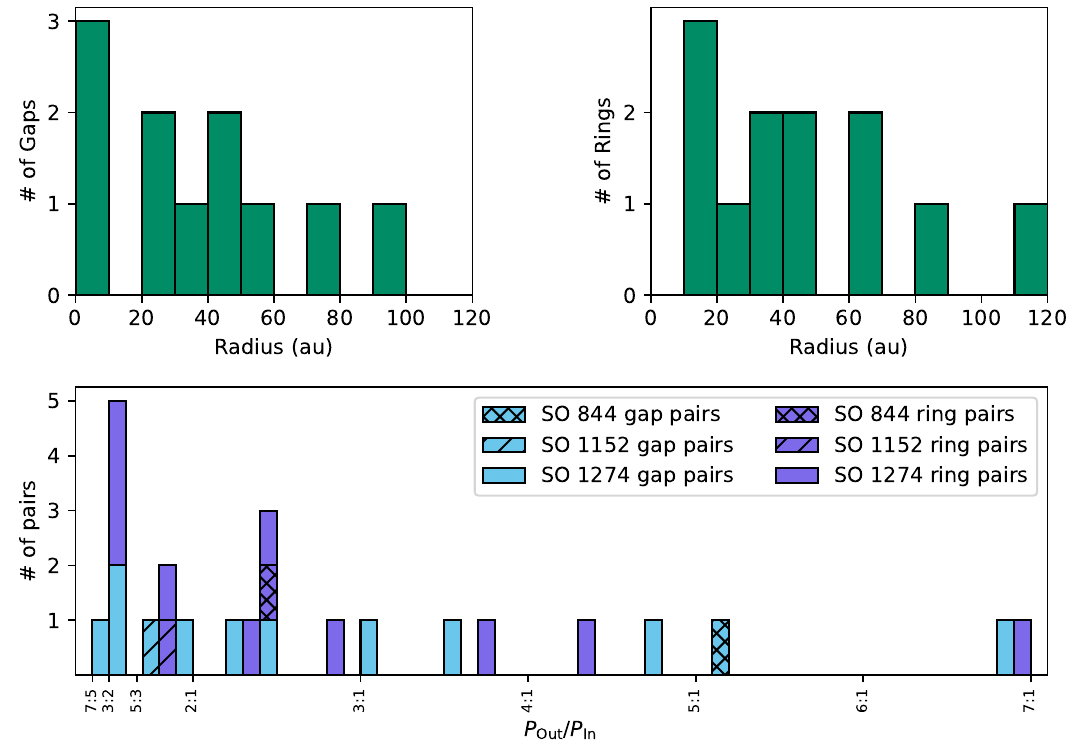}
\end{center}
\caption{Top left: Radial distribution of all gaps identified in the $\sigma$ Orionis sample. Top right: Radial distribution of all rings. Bottom: A stacked histogram of period ratios for pairs of gaps and pairs of rings in disks with more than one gap/ring.  \label{fig:substructurehistograms}}
\end{figure*}

Histograms of the radial locations of the gaps and rings are shown in Figure \ref{fig:substructurehistograms}. The locations have a wide spread, from $<10$ au to $>100$ au, although SO 1274 accounts for all the gaps and rings identified outside 62 au. In the DSHARP survey of disks in nearby star-forming regions, the distribution of detected disk substructures peaks at radii of $\sim30-40$ au \citep{2018ApJ...869L..42H}, whereas the $\sigma$ Orionis distributions peak at $r<20$ au. One possible reason for this difference is that the disks in our sample are on average smaller than those targeted by DSHARP. For the combined sample of disks from DSHARP and the Taurus survey \citep{2018ApJ...869...17L, 2019ApJ...882...49L}, \citet{2023ApJ...952..108Z} found that the peak of the distribution of substructures occurred at smaller radii for disks with effective radii less than 50 au compared to those greater than 50 au. However, these various studies have been performed at different resolutions and used different analysis techniques, so more disk observations as well as a more homogeneous analysis will be required to understand what factors affect the radii at which substructures are most prevalent.

Figure \ref{fig:substructurehistograms} also plots the period ratios of all combinations of gap pairs and ring pairs in the disks with multiple gaps and rings (SO 844, SO 1152, SO 1274), under the assumption that the disk mass is negligible compared to the stellar mass. (\citet{2015ApJ...805..100T} estimated that the actual resonance locations in a disk are offset from simple integer period ratios by $\sim\frac{M_\text{disk}}{M_\ast}$). Non-adjacent pairs are included because it may be the case that a non-planet-related substructure occurs between two planet-related substructures. The SO 844 and SO 1152 substructure pairs are not generally near low-order mean motion resonances. However, SO 1274 presents a more interesting case. Its inner three gaps  (D57, D42, and D27) are close to a 3:2:1 period ratio (more precisely, 3.08:1.95:1). Meanwhile, D76:D57, B66:B50, B88:B66, and D97:D76 are close to 3:2 period ratios (1.53, 1.51, 1.56, and 1.46, respectively). In other words, the five gaps of SO 1274 appear to be arranged in a nearly resonant chain (3:2, 3:2, 3:2, 2:1 from outermost to innermost gap pair), as are B88:B66:B50 (3:2, 3:2). 

\subsubsection{Optical Depths}
We then estimated the optical depths using the expression 
\begin{equation}
\tau_\nu(r) = -\ln\left(1-\frac{I_\nu(r)}{B_\nu(T_d(r))}\right), 
\end{equation}
where $T_d$ is the dust temperature. This expression neglects scattering, which may lead to underestimates of the optical depth. However, use of this expression allows for direct comparison with estimates made for other disks. Following \citet{2018ApJ...869L..46D}, we approximate $T_d$ with 
\begin{equation}\label{eq:temperature}
T_d(r) = \left( \frac{\varphi L_\ast}{8\pi r^2\sigma_\text{SB}}\right)^{0.25},
\end{equation}
where $\varphi = 0.02$ is the chosen flaring angle. This expression assumes that the disk is heated through irradiation from its stellar host, but a nearby massive star can also contribute to heating the outer regions of a disk. Through radiative transfer modeling, \citet{2021MNRAS.503.4172H} found that the midplane temperature of a disk at a separation of 1 pc from a $\theta^1$ Ori C-like system begins to diverage significantly from that of an isolated disk at radii beyond $\sim20$ au. $\theta^1$ Ori C1 has a luminosity of 204,000 $L_\odot$ \citep{2006AA...448..351S}, compared to 41,700 $L_\odot$ for $\sigma$ Ori Aa \citep{2015ApJ...799..169S}, so the impact of external heating on the $\sigma$ Orionis disks should be weaker. In addition, the larger disks in our sample ($R_\text{90}>45$ au, see Section \ref{sec:sizeluminosity}), have relatively large projected separations from $\sigma$ Ori ($\geq 1.7$ pc). 

\begin{figure*}
\begin{center}
\includegraphics{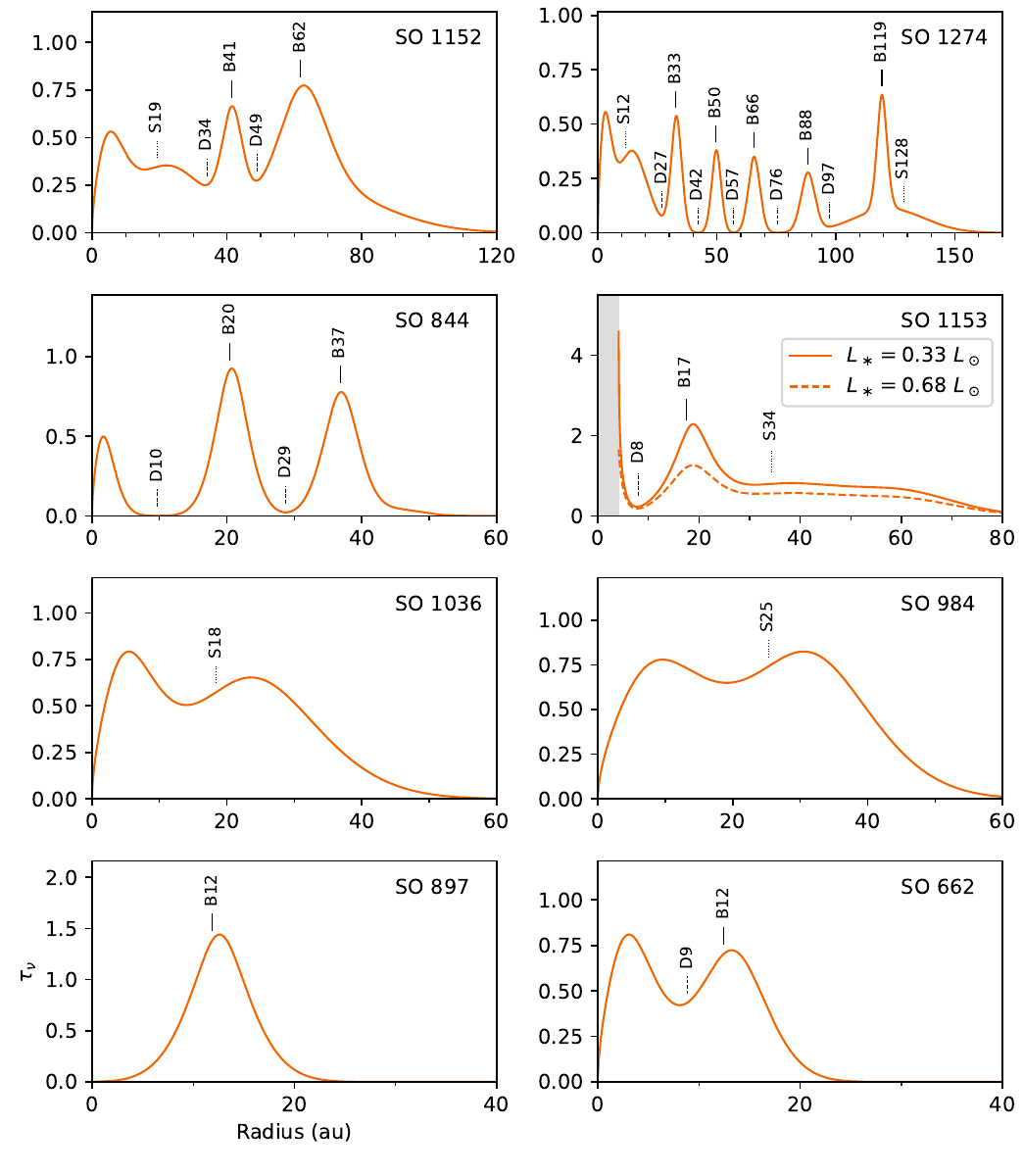}
\end{center}
\caption{Plots of the disk optical depth as a function of radius. Substructures identified from the model radial intensity profiles are labeled. The gray bar in the SO 1153 plot shows where an optical depth cannot be derived because the estimated dust temperature is less than the brightness temperature, indicating that the dust temperatures for this source are underestimated.  \label{fig:opticaldepths}}
\end{figure*}

The optical depths are plotted in Figure \ref{fig:opticaldepths}. In general, the profiles dip within the inner disk due to beam dilution (although SO 897, of course, has a cavity). Some of the shoulders identified in the radial intensity profiles manifest as gap-ring pairs in the optical depth profiles. In most disks, the peak optical depths of the rings and shoulders range from $\sim0.25-0.8$, comparable to the values found for disks in nearby regions \citep[e.g.,][]{2018ApJ...869L..46D, 2020ApJ...891...48H, 2020AA...639A.121F}. \citet{2019ApJ...884L...5S} suggested that the apparent tendency for ring optical depths to fall in this range is a consequence of ongoing planet formation, while \citet{2019ApJ...877L..18Z} found that optically thick rings with high albedo dust grains (which lead to significant scattering) could yield apparent optical depths of $\sim0.6$. 

However, the SO 897 and SO 1153 disks appear to have anomalously high estimated optical depths. Equation \ref{eq:temperature} likely underestimates the temperature at SO 897's ring because depletion of dust inside the cavity would result in strong irradiation of the cavity wall. On the other hand, the anomalously high optical depths of the SO 1153 disk are likely due at least in part to the estimated $L_\ast$ being too low. \citet{2023AA...679A..82M} estimated that $L_\ast=0.33$ $L_\odot$ for SO 1153, which is lower than that of the other sources even though its star is more massive. \citet{2023AA...679A..82M} commented that the spectral type and therefore the $L_\ast$ value for SO 1153 is challenging to derive due to veiling. If the optical depths are recomputed using the median $L_\ast$ of the other sources (0.68 $L_\odot$), they fall more in line with the rest of the sample. In addition, if SO 1153 is embedded, which its classification by \citet{2007ApJ...662.1067H} as a Class I YSO would imply, then Equation \ref{eq:temperature} may not be appropriate since Class I disks tend to be warmer than Class II disks \citep[e.g.,][]{2020ApJ...901..166V}. 

\subsection{Disk fluxes and sizes\label{sec:fluxesandsizes}}
For all disks except SO 897, we used the model radial intensity profiles to calculate the flux with the following equation:

\begin{equation} \label{eq:fluxintegral}
\frac{\cos i}{d^2}\int_0^{r'} 2\pi r I(r) dr,
\end{equation}
where $d$ is the distance to the source and $r'$ is chosen to be some large value (usually a few hundred au) such that the flux integral has reached its asymptotic value. Because SO 897 was not modelled as a flat disk, we measure its flux instead from the model images of the projected disk (although the value is similar to that derived from using Equation \ref{eq:fluxintegral}.) For SO 662, SO 844, and SO 984, our derived flux values differ from those in \citet{2017AJ....153..240A} and \citet{2023AA...679A..82M} by $\sim40-70\%$. This discrepancy arises because they fit point source models to their lower-resolution data. 

We then measured the effective radii using the metrics from \citet{2017ApJ...845...44T}. $R_\text{68}$ is defined as the radius that encloses $68\%$ of the total flux, while $R_\text{90}$ is the radius that encloses $90\%$. Table \ref{tab:disksizesandfluxes} lists the median fluxes and effective radii as well as their $68\%$ confidence intervals. For all disks, including SO 897, the flux percentages are computed with respect to the flux calculated with Equation \ref{eq:fluxintegral}.

\begin{deluxetable}{lccc}
\tablecaption {Disk fluxes and sizes \label{tab:disksizesandfluxes}}
\tablehead{\colhead{Source} &\colhead{Flux\tablenotemark{a}}&\colhead{$R_\text{68}$\tablenotemark{b}} & \colhead{$R_\text{90}$\tablenotemark{b}}\\
\colhead{} &\colhead{(mJy)}&\colhead{(au)} & \colhead{(au)}}
\startdata
SO 662 &$2.09\pm0.03$ &$14.2\pm0.2$&$17.7\pm0.3$\\
SO 844 &$4.95\pm0.05$ &$36.5\pm0.2$&$40.4\pm0.3$\\
SO 897 & $1.99\pm0.02$&$14.8\pm0.2$&$17.9\pm0.2$\\
SO 984 &$8.46\substack{+0.09 \\ -0.08}$&$34.3\pm0.2$&$43.7\pm0.3$\\
SO 1036 &$6.34\pm0.06$&$28.9\pm0.2$&$38.1\substack{+0.3\\-0.4}$ \\
SO 1152 &$9.32\pm0.06$&$65.3\pm0.3$&$81.9\pm0.6$\\ 
SO 1153 & $16.1\pm0.1$&$52.0\pm0.2$&$67.7\pm0.3$\\
SO 1274 & $15.5\pm0.2$&$113.2\substack{+0.7 \\ -0.8}$&$126.7\substack{+0.7 \\ -0.6}$ \\
\enddata 
\tablenotetext{a}{Error bars do not include $\sim10\%$ systematic flux calibration uncertainty}
\tablenotetext{b}{Error bars do not include the uncertainty in distance.}
\end{deluxetable}

In models of disks with dynamic pressure bumps, \citet{2022AA...668A.104S} found that the outermost bump was generally located near $R_\text{68}$. For SO 1152, SO 1274, SO 844, SO 897, and SO 662, the radial location of the outermost identified ring is within several au of the disk's $R_\text{68}$ value, providing some support to the idea that disk sizes are often controlled by the locations of their pressure bumps. For SO 1153, the $R_\text{68}$ value of 52 au is well outside the ring detected at 18 au or the shoulder detected at 34 au, but it is possible that the extended emission tail at larger radii may harbor unresolved rings. Alternatively, given that \citet{2007ApJ...662.1067H} categorized SO 1153 as a Class I YSO, the relationship between disk size and pressure bump location may differ for Class I and II disks. 

\subsection{Size-luminosity relationship\label{sec:sizeluminosity}}

Several observational studies have found a correlation between disk sizes and luminosities, although the scalings vary between different star-forming regions \citep[e.g.,][]{2017ApJ...845...44T, 2018ApJ...865..157A, 2020ApJ...895..126H}. Following these previous works, the disk luminosity $L_\text{mm}$ is defined as the disk flux rescaled to a distance of 140 pc. We first sampled flux values from a normal distribution with a mean and standard deviation corresponding to the values in Table \ref{tab:disksizesandfluxes}. To account for the systematic flux calibration uncertainty, we then multiplied the flux samples by scaling factors randomly drawn from a normal distribution with a mean of 1 and standard deviation of 0.1. To account for the uncertainty in distance, we generated posterior samples for the distances using the Interactive Distance Estimation tool\footnote{\url{https://github.com/ElisaHaas25/Interactive-Distance-Estimation/tree/main}}, which implements the method described in \citet{2021AJ....161..147B} to estimate distances from the \textit{Gaia} catalog \citep{2021AA...649A...1G}. The flux values were then rescaled using randomly drawn samples from the distance posteriors. Similarly, the distribution of $R_\text{68}$ values was generated by sampling from a normal distribution with a mean and standard deviation corresponding to the values in Table \ref{tab:disksizesandfluxes} and rescaling with randomly drawn distance values. We calculated the medians of these $L_\text{mm}$ and $R_\text{68}$ distributions as point estimates and the 16th and 84th percentiles to obtain $1\sigma$ uncertainties. 

\begin{figure}
\begin{center}
\includegraphics{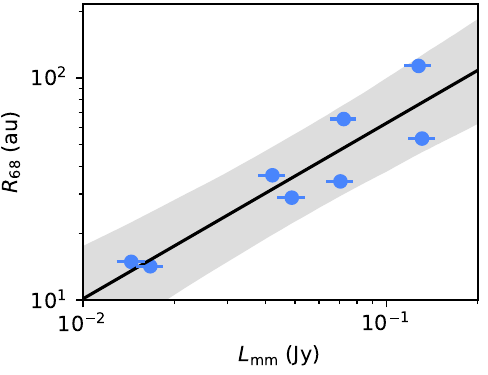}
\end{center}
\caption{A plot of $R_\text{68}$ vs. $L_\text{mm}$ for the observed $\sigma$ Orionis disks. Following \citet{2017ApJ...845...44T}, $L_\text{mm}$ is defined as the flux scaled to a distance of 140 pc. The black line corresponds to the median parameters from the linear fit, while the gray shaded region shows the $68\%$ confidence interval. (Note that the error bars on $R_\text{68}$ are too small to be visible.) \label{fig:sizeluminosity}}
\end{figure}

As in the aforementioned works, we model the relationship between $R_\text{68}$ and $L_\text{mm}$ with the following equation:

\begin{equation}
\log \left(R_\text{68}/\text{au} \right)= \alpha + \beta \log \left(L_\text{mm}/\text{Jy}\right). 
\end{equation}
The scatter of $\log R_\text{68}$ about the regression line is assumed to be Gaussian with standard deviation $\sigma$. The data are fit using a Python port\footnote{\url{https://github.com/jmeyers314/linmix}} of \texttt{linmix}, a Bayesian linear fitting code by \citet{2007ApJ...665.1489K} that accounts for uncertainties in both the independent and dependent variables. The median and $1\sigma$ uncertainties of the posterior distributions are reported in Table \ref{tab:sizeluminosity}. The linear fit is plotted with the datapoints in Figure \ref{fig:sizeluminosity}. 

\begin{deluxetable}{cc}
\tablecaption {Size-luminosity regression results \label{tab:sizeluminosity}}
\tablehead{\colhead{Parameter} &\colhead{Value}}
\startdata
$\alpha$ & $2.6\pm0.3$\\
$\beta$ & $0.8\pm0.2$ \\
$\sigma$ & $0.17\substack{+0.10\\-0.05}$\\
$\hat\rho$ (Correlation coefficient) & $0.91\substack{+0.06\\-0.15}$
\enddata  
\end{deluxetable}

As in nearby star-forming regions \citep[e.g.,][]{2017ApJ...845...44T, 2018ApJ...865..157A, 2020ApJ...895..126H}, we find a strong correlation between $\log L_\text{mm}$ and $\log R_\text{68}$, with a linear correlation coefficient of 0.91. Our slope, $\beta = 0.8\pm0.2$, is slightly steeper than the values derived for various nearby regions ($0.22-0.6$), but consistent within the uncertainties. Likewise, our intercept, $\alpha = 2.6\pm0.3$, is somewhat higher than the values for nearby regions ($1.7-2.2$), but also consistent within uncertainties. Observing additional disks in $\sigma$ Orionis and measuring disk sizes from higher resolution observations in other regions will be needed to assess whether there is a genuine difference in the size-luminosity relationship. On the other hand, \citet{2018ApJ...860...77E} and \citet{2021ApJ...923..221O} found weak or no evidence for a correlation in the ONC and a combined ONC/OMC1 sample, respectively. The former measured a slope of $0.09\pm0.07$ and the latter measured $0.17\pm0.05$. These works observed disks with projected separations within a few tenths of a pc from $\theta^1$ Ori C, so the different size-luminosity relationships in the ONC and OMC1 compared to our $\sigma$ Orionis sample may be a reflection of the more extreme effects of external photoevaporation. However, the presence of intracluster material in the ONC and OMC1 and the accompanying $uv$ cut employed by \citet{2018ApJ...860...77E} and \citet{2021ApJ...923..221O} may also have introduced greater uncertainty. 

\citet{2018ApJ...865..157A} tentatively identified $R_\text{68}-L_\ast$ and $R_\text{68}-M_\ast$ correlations. Because we selected targets in a relatively narrow $M_\ast$ range (and therefore in a relatively narrow $L_\ast$ range), we do not have the dynamic range to test for correlations.

\subsection{Dust trapping analysis}

Grain size estimates derived from multi-frequency observations and measurements of gas pressure profiles suggest that millimeter continuum rings are often dust traps \citep[e.g.,][]{2020ApJ...898...36L, 2020MNRAS.495..173R, 2021AA...648A..33M, 2021ApJS..257...14S}. In the absence of these kinds of data, \citet{2018ApJ...869L..46D} argued that dust rings with smaller widths $w_d$ than the pressure scale height $h_p$ were likely to be dust traps because a gas pressure bump that is narrower than $h_p$ would not be stable (however, a dust ring being wider than $h_p$ does not necessarily imply that it is \textit{not} a dust trap). 

\begin{deluxetable}{lccccc}
\tablecaption {Estimated properties of high-contrast rings \label{tab:highcontrastrings}}
\tablehead{\colhead{Source} &\colhead{Ring ID}&\colhead{$T_\text{dust}$\tablenotemark{a}} & \colhead{$w_d$}&\colhead{$h_p$}&\colhead{$w_d/h_p$}\\
\colhead{} &\colhead{}&\colhead{(K)} & \colhead{au}&\colhead{(au)}}
\startdata
SO 844 & B20 &24.4&2.8&1.4 &2.0\\
& B37 &18.2&3.0&2.8&1.1\\
\hline
SO 897 & B12 & 34.6 & 3.9& 0.6 &6.5\\
\hline
SO 1153 & B17 & 27.0  & 5.2&0.8 & 6.5 \\
\hline
SO 1274 & B33 &19.7& 2.4&2.1&1.2\\
& B50 & 16.0 & 2.1&3.5&0.6\\
& B66 &14.0 & 2.6&4.9&0.5\\
& B88 &12.0 & 3.1&7.1 &0.4\\
&B119 & 10.3 & 2.3&10.4&0.2\\
\enddata  
\tablenotetext{a}{$T_\text{d}$ is calculated with $L=0.68$ $L_\odot$ for SO 1153 and with the $L_\ast$ values in Table \ref{tab:hoststars} for the other disks.}
\end{deluxetable}

To assess whether the rings we detect in $\sigma$ Orionis might be dust traps, we calculate the ratio of $w_d$ to $h_p$ for the high-contrast rings (those with depth values $<0.5$ in Table \ref{tab:substructureproperties}). The pressure scale height is given by 
\begin{equation}
h_p(r) = \sqrt{\frac{k_B T_d(r) r^3}{\mu m_p G M_\ast}},
\end{equation}
where $T_d(r)$ is estimated using equation \ref{eq:temperature} and $\mu=2.37$ is the mean molecular weight.
We set $w_d$ equal to the value of $\sigma_i$ inferred for the corresponding ring in the parametric model (see Table \ref{tab:modelparameters}). (Note that this is different from the ring widths quoted in Table \ref{tab:substructureproperties}, which would be equivalent to the FWHM of a single Gaussian ring.) The estimates for $T_\text{dust}$, $w_d$, $h_p$, and $w_d/h_p$ are listed in Table \ref{tab:highcontrastrings}. 

Provided that the rings are resolved, the dominant source of uncertainty in the calculation of $w_d/h_p$ is likely the dust temperature estimate for $h_p$. While it is not straightforward to quantify the temperature uncertainty, one can obtain some idea by comparing the dust temperature estimates from \citet{2018ApJ...869L..46D} using equation \ref{eq:temperature} to those derived either from thermochemical modeling \citep{2021ApJS..257....5Z, 2021ApJS..257...14S} or from multi-frequency continuum modeling \citep{2024ApJ...971..129C} for the Elias 24, HD 163296, and AS 209 disks. In these cases, the empirically derived temperatures at the rings were within $\sim50\%$ of the estimates from the analytic model. $h_p$ also depends on $M_\ast$, the values of which were derived by \citet{2023AA...679A..82M} using the \citet{2015AA...577A..42B} evolutionary models for SO 844 and the \citet{2016AA...593A..99F} magnetic models for SO 897, SO 1153, and SO 1274. For the relevant stellar mass ranges, \citet{2021ApJ...908...46B} found that the mass estimates from evolutionary models usually agreed with dynamical mass estimates within $5-10\%$, so the uncertainty in $T_d$ dominates over the uncertainty in $M_\ast$. Thus, we estimate an uncertainty on $h_p$ of $\sim25\%$.

In the SO 1274 disk, $w_d/h_p<1$ for B50, B66, B88, and B119. Taking into account our rough estimate for the uncertainty of $h_p$, we consider these rings likely to be dust traps. Strong dust trapping may be responsible for maintaining the large size of the SO 1274 disk ($R_\text{90} = 127$ au). For the other high-contrast rings identified, $w_d/h_p>1$. However, they are all located at smaller radii than the SO 1274 rings, and thus the scale heights at those locations would likely be smaller. The larger $w_d/h_p>1$ values may be a consequence of the identified rings being under-resolved, or possibly even being composed of multiple rings, as seen in disks such as HD 169142 \citep{2019AJ....158...15P}. The values of $w_d/h_p$ are slightly larger than 1 for B37 in SO 844 and B33 in SO 1274, but given the large uncertainties in $h_p$, they warrant further examination with improved dust temperature estimates and higher resolution observations.

\subsection{Planet mass estimates}
If the disk gaps are due to planets, then one can use the gap widths to estimate the masses of the embedded planets. We use the fitting relations introduced in \citet{2018ApJ...869L..47Z}, which are implemented in code presented in \citet{2022MNRAS.510.4473Z}. In brief, we assume that each observed gap corresponds to one protoplanet. The semi-major axis $r_p$ of the protoplanet is set to the radial location of its gap (from Table \ref{tab:substructureproperties}). The fractional gap width $\Delta$ is measured from the model radial intensity profile. The average dust surface density $\Sigma_\text{dust, avg}$ is estimated from the model optical depth profile between 1.1$r_p$ and $\min(2r_p, R_\text{90})$ using $\Sigma_\text{dust}(r)= \tau_\nu(r)/\kappa_{\nu,\text{abs}}$. (For SO 1153, we assume that $L_\ast=0.68$ $L_\odot$). Following \citet{2018ApJ...869L..47Z}, the dust absorption opacity is set to 0.43 cm$^{2}$ g$^{-1}$, corresponding to the standard DSHARP dust composition \citep{2018ApJ...869L..45B} with a maximum grain size of $a_\text{max}=0.1$ mm and a size distribution of $n(a)\propto a^{-3.5}$ (\citet{2018ApJ...869L..47Z} refer to this as the "DSD1" distribution). Given a value of $\Sigma_\text{dust, avg}$, the maximum Stokes number St$_\text{max}$ is estimated from the grid of hydrodynamical simulations in \citet{2018ApJ...869L..47Z} for the DSD1 distribution and the aspect ratio $h_p/r$ (0.05, 0.07, and 1.0) closest to the value estimated at the gap. However, if St$_\text{max}$ corresponds to a gas surface density that exceeds the limit for gravitational stability (Toomre $Q=1$, with the thermal profile from Equation \ref{eq:temperature}), then the next largest value of St$_\text{max}$ from the model grid is used. Then, a planet mass $M_p$ can be calculated from the scaling relations in \citet{2018ApJ...869L..47Z} for a given $\Delta$, St$_\text{max}$, $\frac{h_p}{r}$ and viscosity parameter $\alpha$. We assume that $\alpha=10^{-3}$. All other things held equal, changing $\alpha$ by an order of magnitude changes the $M_p$ estimate by a factor of $\sim2$.


\begin{deluxetable*}{ccccccccc}
\tablecaption {Estimated planet masses\label{tab:planetmasses}}
\tablehead{\colhead{Source} &\colhead{Gap ID}&\colhead{$\Delta$} & \colhead{$\Sigma_\text{dust, avg}$}&\colhead{$h_p/r$}&\colhead{$\Sigma_{\text{gas}, Q=1}$}&\colhead{St$_\text{max, used}$}&\colhead{$M_p$}&\colhead{Uncertainty}\\&&&\colhead{(g cm$^{-2}$)}& & \colhead{(g cm$^{-2}$)}&\colhead{($\times 5.23\text{e}-4$)}&\colhead{($M_\text{Jup}$)}&\colhead{($\log M_p$)}
}
\startdata
SO 662 & D9 & 0.29 & 1.3 & 0.05 & 1044 & 0.3 & 0.46 & $\substack{+0.13\\-0.16}$ \\
\hline
SO 844 & D10 & 0.74 & 0.53 & 0.06 & 736 & 0.3 & 7.9 & $\substack{+0.13\\-0.16}$ \\
& D29 & 0.26 & 1.2 & 0.07 & 110 & 0.3 & 0.23 & $\substack{+0.13\\-0.16}$\\
\hline
SO 1152 & D34 & 0.29 & 1.3 & 0.07 & 94 & 1 & 0.36 &$\substack{+0.16\\-0.14}$ \\
& D49 & 0.17 & 1.2 & 0.07 & 50 & 1 & 0.04 & $\substack{+0.16\\-0.14}$\\
\hline
SO 1153 & D8 & 0.5 & 1.4 & 0.04 & 1490 & 0.3 & 3.9 &  $\substack{+0.13\\-0.16}$\\
\hline
SO 1274 & D27 & 0.30 &0.43 &0.06&151&0.3 & 0.52 & $\substack{+0.13\\-0.16}$\\
 & D42 & 0.22 & 0.23 &0.07& 69 & 1 &0.13&$\substack{+0.16\\-0.14}$\\
 & D57 & 0.16 & 0.22 & 0.07&40.5 & 1 & 0.035 & $\substack{+0.16\\-0.14}$ \\
 & D76 & 0.18 & 0.40 & 0.08&24.7 & 3 & 0.07 & $\substack{+0.14\\-0.17}$ \\
 & D97 & 0.22 & 0.56&0.08&16&3&0.17&$\substack{+0.14\\-0.17}$
\enddata  
\end{deluxetable*}

\begin{figure*}
\begin{center}
\includegraphics{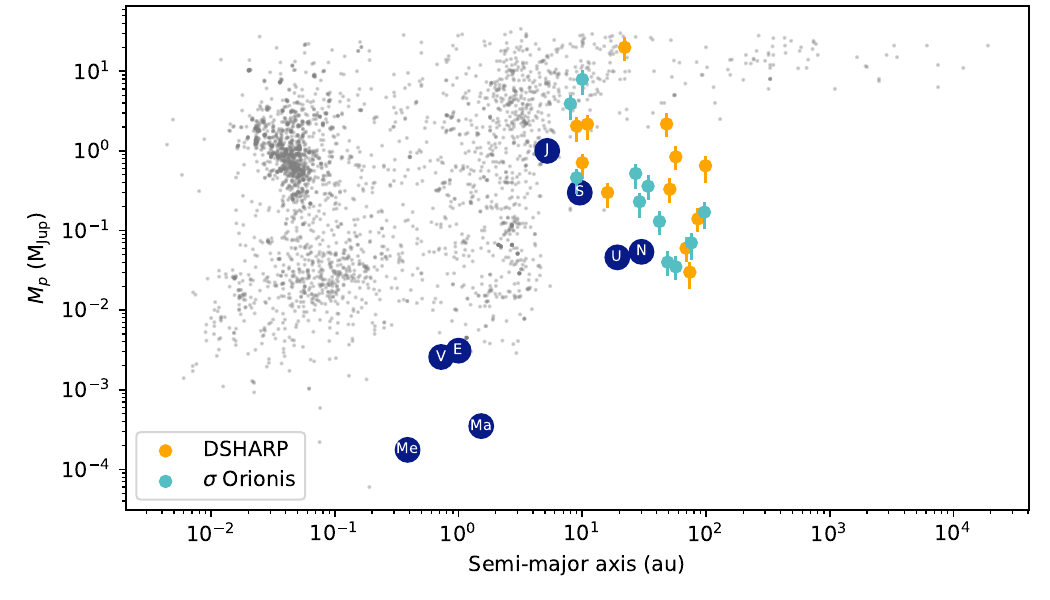}
\end{center}
\caption{A comparison of the planet masses inferred in the $\sigma$ Orionis sample in this work (turquoise dots) and those inferred from well-resolved disk gaps in the DSHARP survey of nearby star-forming regions (orange dots) \citep{2018ApJ...869L..47Z}. The DSHARP values correspond to $\alpha=10^{-3}$ and the DSD1 dust distribution. The error bars show the uncertainties calculated by the linear fitting method. The gray dots correspond to confirmed exoplanets (retrieved from \url{https://exoplanetarchive.ipac.caltech.edu/} on 10 July 2024). The Solar System planets are plotted in dark blue. \label{fig:planetmasses}}
\end{figure*}

The planet mass estimates are listed in Table \ref{tab:planetmasses}. The estimates are compared for those of the DSHARP survey, which targeted nearby star-forming regions, in Figure \ref{fig:planetmasses}. The overall ranges of inferred planet masses and semi-major axes are similar, although the $\sigma$ Orionis values tend to be concentrated more at somewhat smaller masses and semi-major axes. 

The planet mass uncertainties listed in Table \ref{tab:planetmasses} and plotted in Figure \ref{fig:planetmasses} are the formal errors calculated from the \citet{2018ApJ...869L..47Z} linear fitting method, but the process of inferring planet masses from disk gap properties has other sources of uncertainty. Planet mass estimates are sensitive to the disk temperature, gas surface density, and grain size distribution, the values of which are crudely approximated. Furthermore, while our calculations assumed that every gap has a planet, simulations have shown that in low-viscosity disks, a single planet may be responsible for multiple gaps \citep[e.g.,][]{2017ApJ...850..201B, 2017ApJ...843..127D, 2018ApJ...869L..47Z}. In addition, other studies have illustrated that including effects such as radiative cooling, self-gravity, magnetized disk winds, or migration in simulations of planet-disk interactions can affect the morphology of substructures created by a given planet \citep[e.g.,][]{2019ApJ...878L...9M, 2019MNRAS.482.3678M, 2020MNRAS.493.2287Z, 2023ApJ...946....5A}.  

\section{Discussion\label{sec:discussion}}
\subsection{Implications for disk evolution and planet formation in $\sigma$ Orionis}

Our observations suggest that disk substructures are able to form and survive across a range of external UV environments, from a few $G_0$ to $10^3$ $G_0$. Substructures are detected even in very compact disks ($R_\text{68}<15$ au) in $\sigma$ Orionis. However, multi-ringed systems are only detected at $d_\text{proj}>2.1$ pc. This may be a matter of resolution, since larger disks tend to be located at greater separations. If the substructures are due to mechanisms such as zonal flows or planet-disk interactions, they should be more readily detectable at larger radii because the scale height sets their characteristic width, and the scale height increases with radius \citep[e.g.,][]{1999ApJ...514..344B,2009ApJ...697.1269J}. Nevertheless, SO 844 presents an interesting contrast with SO 984 and SO 1036. Despite having similar millimeter continuum disk sizes, SO 844 ($d_\text{proj}=2.1$ pc) has two deep, wide gaps, while SO 1036 and SO 984 ($d_\text{proj}=1.2$ pc for both) do not have clearly detected disk gaps (although they each have shoulders).   

If the detected substructures are due to planet-disk interactions, their widths and locations suggest that ice and gas giants can form in $\sigma$ Orionis disks on Solar System scales. However, whereas more than half of the DSHARP systems have gaps detected outside 50 au \citep{2018ApJ...869L..42H}, SO 1274 is the only one of the eight disks in our $\sigma$ Orionis survey that does (although SO 1152 has a gap at 49 au). Of course, neither sample was chosen in an unbiased manner, but we can use our derived size-luminosity relation (Section \ref{sec:sizeluminosity}) to estimate how many other disks in $\sigma$ Orionis might have millimeter continuum gaps outside 50 au. We assume that $R_\text{68}$ must be greater than 50 au in order for a disk to have a continuum gap outside 50 au (which holds for our sample and is expected based on the \citet{2022AA...668A.104S} models of dynamical pressure bumps). Our size-luminosity relation predicts that a disk of this size has a 1.3 mm flux of 9 mJy at a distance of 402 pc. Of the disks not in our sample, only SO 540 meets this threshold \citep{2017AJ....153..240A}. The extent of the millimeter continuum emission, though, does not necessarily denote an upper bound on where protoplanets might be present. Gaps in scattered light and molecular line emission, as well as velocity ``kinks,'' have been detected outside the millimeter continuum in some disks and in some cases have been hypothesized to be due to planet-disk interactions \citep[e.g.,][]{2018ApJ...860L..13P, 2018ApJ...863...44A, 2021ApJS..257....3L}. 

The apparent preponderance of substructure pairs near 3:2 period ratios in the SO 1274 disk is intriguing given that the period ratios of the \textit{Kepler} multiplanet systems show a peak near 3:2 as well \citep{2014ApJ...790..146F,2015MNRAS.448.1956S}. The \textit{Kepler} planets have relatively small periods (generally less than a couple hundred days), so their architectures cannot be compared directly with disks imaged by ALMA. Nevertheless, SO 1274 is potentially a useful system to model to understand how multiplanet systems in resonant configurations arise.

\citet{2020MNRAS.492.1279S} modelled the effects of external photoevaporation on disks with smooth surface density profiles and found that disks exposed to stronger external FUV radiation fields experienced more rapid dust depletion via radial drift, thereby shortening their lifetimes. However, \citet{2024AA...681A..84G} found that including substructures in the models could prolong dust disk lifetimes to a few Myr for external FUV radiation levels up to $10^3$ $G_0$. The detection of substructures in our disk targets, which have an estimated age of $3-5$ Myr \citep{2002AA...384..937Z, 2002AA...382L..22O, 2004MNRAS.347.1327O, 2019AA...629A.114C} and external FUV fields of $10^2-10^3$ $G_0$ \citep{2023AA...679A..82M}, supports the need to consider the role of substructures when modelling the impact of external photoevaporation on disk evolution. 

The observed correlation between disk size and luminosity in nearby star-forming regions has been hypothesized to be due to dust trapping by substructures \citep[][]{2017ApJ...845...44T,2024AA...688A..81D}. The prevalence of substructures in our $\sigma$ Orionis sample and its strong size-luminosity correlation broadly fits within this picture. The absence of a clear size-luminosity correlation in the ONC \citep{2018ApJ...860...77E, 2021ApJ...923..221O} may be a consequence of the higher external UV radiation fields ($\geq10^4$ $G_0$, \citet{2010ApJ...725..430M}) either destroying substructures or inhibiting their formation in the first place. 

$\sigma$ Orionis seems to be similar to other star-forming regions observed so far in that (approximately) axisymmetric gaps and rings are the dominant forms of substructures detected. However, it is worth commenting on the kinds of substructures that have been observed in other star-forming regions \citep[][]{2018ApJ...869L..41A, 2020ARAA..58..483A}, but do not appear in our sample. No large cavities ($r>20$ au), spiral arms, or vortex-like crescents are visible in our data (although SO 897 features a small cavity). Besides SO 897, eight additional members of $\sigma$ Ori (SO 299, SO 411, SO 540, SO 587, SO 818, SO 908, SO 1267, and SO 1268) have been classified as transition disks or transition disk candidates based on their infrared SEDs \citep[][]{2007ApJ...662.1067H, 2016ApJ...829...38M}. While most of these disks are fainter than the ones we observed, SO 411 and SO 540 have relatively large 1.3 mm fluxes (5.2 and 10.7 mJy, respectively, as measured by \citet{2017AJ....153..240A}). \citet{2018ApJ...869L..41A} found that transition disks and non-transition disks followed similar size-luminosity relationships in nearby-star forming regions. If the same holds true in $\sigma$ Ori, then it would be worthwhile to image SO 411 and SO 540 to determine whether they have large cavities. Meanwhile, spiral arms that have been detected in millimeter continuum observations of disks in nearby star-forming regions have generally been much lower-contrast than rings \citep[e.g.,][]{2016Sci...353.1519P, 2018ApJ...860..124D, 2018ApJ...869L..43H, 2018ApJ...869L..44K}. Thus, we would expect similar spiral arms to be challenging to detect in the $\sigma$ Orionis disks. Vortex-like crescents have thus far disproportionately been detected in disks hosted by stars greater than solar mass \citep[][]{2023ASPC..534..423B}, whereas all of our targets are roughly solar mass or below. 

\subsection{Non-planetary explanations for $\sigma$ Orionis substructures}
While the substructures detected are plausibly explained by planet-disk interactions, a number of models have been presented to explain substructures without invoking planet-disk interactions. For recent reviews of substructure formation models, see \citet{2023ASPC..534..423B} and \citet{2023ASPC..534..465L}. We comment briefly on some commonly studied alternatives. 

One of the most popular alternative explanations has been that the substructures result from dust evolution at or near molecular snowlines \citep[e.g.,][]{2015ApJ...806L...7Z, 2016ApJ...821...82O, 2017ApJ...845...68P}. Several studies have argued that snowlines are unlikely to account for the majority of substructures, given that they do not systematically appear at the estimated radial locations of snowlines \citep[][]{2018ApJ...869L..42H, 2018ApJ...869...17L, 2018ApJ...867L..14V}. That said, the locations of molecular snowlines are often contested even in systems that have been much more extensively observed and modelled \citep[e.g.,][]{2013Sci...341..630Q, 2016Natur.535..258C, 2017NatAs...1E.130Z, 2017AA...599A.101V, 2018ApJ...864L..23V}. The models of substructure formation via snowlines alone, though, have not yielded a simple way to reproduce the heterogeneity in the number of gaps and rings observed in different disks. \citet{2024MNRAS.527.2049C} suggested that planet-disk interactions could shift snowline locations by altering the disk thermal structure, and thus a combination of planet-disk interactions and snowlines could explain the lack of an obvious pattern in the substructure numbers or locations. However, there is debate over whether molecular snowlines can create the high-contrast dust rings that have been imaged by ALMA. For example, \citet{2017AA...600A.140S} modelled the effect of CO freezeout on dust evolution and found that it results in only a minor decrease in the dust surface density just interior to the CO snowline. 

SO 897 features a disk with a small inner cavity (outer ring peaks at $r\sim12$ au). Models from \citet{2019MNRAS.487..691P} show that cavities of this size can be produced through photoevaporation due to X-rays from the stellar host, but photoevaporation alone would most likely lead to an accretion rate significantly lower than SO 897's measured value of $4.7\times10^{-9}$ $M_\odot$ yr$^{-1}$ \citep{2023AA...679A..82M}. \citet{2021AA...655A..18G} found that models with both dead zones and X-ray photoevaporation could allow high accretion rates to be sustained while opening a disk cavity. A telltale feature of their models is a small disk within the cavity, but our angular resolution is insufficient to determine whether SO 897's cavity has an inner disk. 

Magnetohydrodynamic winds and zonal flows can generate substructures on scales similar to those produced by planet-disk interactions \citep[e.g.,][]{2014ApJ...796...31B, 2018AA...617A.117R, 2018MNRAS.477.1239S,2023MNRAS.523.4883H}. As discussed in \citet{2023ASPC..534..423B}, observationally distinguishing these mechanisms is not straightforward in the absence of a direct detection of a planet. A particular difficulty is that only upper limits so far have been derived for the magnetic field strengths of protoplanetary disks \citep[e.g.,][]{2019AA...624L...7V,2021ApJ...908..141H}.   

In most cases, the effects of the external FUV radiation field have not been considered in substructure formation models. Testing the effects of a range of external FUV levels on substructure development may offer an additional way to discriminate between models.

\subsection{Caveats and future prospects}
The external FUV radiation fields impinging on our targets could be significantly lower than the estimated values listed in Table \ref{tab:hoststars}. First, the true separations from $\sigma$ Ori could be significantly larger than the projected separations. Second, while extinction is currently low in $\sigma$ Orionis, higher amounts of intracluster material in the past may have protected the disks from external FUV radiation \citep[e.g.,][]{2022MNRAS.512.3788Q}. Finally, these disks could have been exposed to lower external FUV radiation in the past if they migrated from the outskirts of their cluster \citep[e.g.,][]{2019MNRAS.490.5478W}. 

\citet{2017AJ....153..240A} and \citet{2023AA...679A..82M} showed that disk millimeter continuum fluxes tend to decrease as projected separations from $\sigma$ Ori decrease. Given that disk sizes and luminosities appear to be well-correlated in this region (Section \ref{sec:sizeluminosity}), this implies that millimeter continuum disk sizes (and thus the radial range over which millimeter continuum substructures can occur) also tend to decrease as projected separations decrease. We noted earlier in the discussion that disks with millimeter continuum gaps beyond a radius of 50 au are likely rare in $\sigma$ Orionis. While external photoevaporation is plausibly responsible for the apparent absence of large disks (and by extension, wide separation millimeter continuum substructures) at smaller projected separations from $\sigma$ Ori, age effects may also contribute. Disks in the 5-11 Myr old Upper Sco region are systematically smaller than those in the 1-3 Myr old Lupus or 2-3 Myr old Cha I regions \citep{2017ApJ...851...85B, 2020ApJ...895..126H}. \citet{2020ApJ...895..126H} suggested that this trend could either be the consequence of older disks having experienced more radial drift or Upper Sco having stronger external UV fields compared to the other regions studied. With an estimated age of $3-5$ Myr \citep{2002AA...384..937Z, 2002AA...382L..22O, 2004MNRAS.347.1327O, 2019AA...629A.114C}, $\sigma$ Orionis appears to be somewhat older than the nearby regions with a significant population of large disks. Observations of the western population of NGC 2024, which has an estimated age of $\sim1$ Myr and shows evidence of mass loss due to external photoevaporation \citep{2020AA...640A..27V}, may help to disentangle the effects of age and external radiation fields. In addition, \citet{2018AJ....156..271L} suggested that $\sigma$ Orionis has smaller disk fractions compared to some other regions that are several Myr old (such as Cha I) due to higher stellar density in $\sigma$ Orionis. Higher stellar densities increase the probability of close stellar encounters, which can lead to the truncation or destruction of disks \citep[e.g.,][]{1993MNRAS.261..190C, 2014AA...565A.130B, 2018MNRAS.478.2700W}. However, models from \citet{2018MNRAS.478.2700W} suggest that for $\sigma$ Orionis-like conditions, external photoevaporation should play a larger role than dynamical truncation in setting disk sizes.

Our observations only included stars between 0.4 and 0.9 $M_\odot$. Models indicate that mass loss due to external photoevaporation is more severe around lower-mass stars \citep[e.g.,][]{2018MNRAS.481..452H}. Counterintuitively, \citet{2023AA...679A..82M} found that disk masses in $\sigma$ Orionis for stars above 0.4 $M_\odot$ showed a trend with projected separation from $\sigma$ Ori, whereas those below 0.4 $M_\odot$ did not. Given that surveys of very low-mass stars in nearby regions are detecting disk substructures \citep{2021AA...645A.139K,2024ApJ...966...59S}, it is plausible that the lower-mass stars in $\sigma$ Orionis also have disks with substructures. However, this needs to be confirmed with observations.  

Some of the observational tests used to probe the origins of substructures in nearby regions, such as direct imaging searches for protoplanet emission \citep[e.g.,][]{2018AA...617A..44K, 2021AJ....161..146J, 2021AA...652A.101A, 2023MNRAS.522L..51H}, molecular line kinematics \citep[e.g.,][]{2018ApJ...860L..13P,2018ApJ...860L..12T,2023ASPC..534..645P}, and resolved spectral index measurements \citep[e.g.,][]{2016ApJ...829L..35T,2019ApJ...883...71C,2020ApJ...891...48H, 2020ApJ...898...36L}, will be difficult or impossible to apply to $\sigma$ Orionis with existing facilities due to the larger distance to the cluster and its relatively small disks. However, such studies may become feasible with Extremely Large Telescope-class facilities, the Next Generation Very Large Array, or the proposed ALMA $\times$ 10 upgrade. 
\section{Summary\label{sec:summary}}
We used ALMA to image the 1.3 mm continuum of eight disks in the $\sigma$ Orionis cluster, producing the highest resolution observations to date of disks in this region. Our key findings are as follows: 
\begin{itemize}
\item Gaps, rings, or cavities are visible in the images of five of the eight disks. Through visibility modeling, we infer that a sixth disk also has a gap and the remaining two have shoulder-like structures. The disks appear to be largely axisymmetric, which is also the case for most disks observed in nearby star-forming regions. 
\item Three of the disks with gaps or cavities have $R_\text{90}$ values less than 50 au, illustrating the diversity of structures in compact disks. 
\item The large SO 1274 disk exhibits an especially rich set of substructures, with at least five deep, wide gaps located from 27 to 97 au. The substructures appear to be arranged nearly in a resonant chain. The outer rings in this system are narrow and likely dust traps. 
\item Disk sizes and luminosities are well-correlated within the sample, a characteristic that has also been observed in nearby star-forming regions. In most of the sample, the outermost detected ring is located near the measured value of $R_\text{68}$, supporting the proposition that substructures play a role in setting apparent dust disk sizes. 

\item Given the disk fluxes measured in moderate-resolution surveys of $\sigma$ Orionis \citep{2017AJ....153..240A,2023AA...679A..82M}, the size-luminosity relationship measured from our high-resolution observations implies that most disks in $\sigma$ Ori are small and thus few disks in the cluster are likely to have millimeter continuum gaps beyond a radius of 50 au, in contrast to the DSHARP sample of disks in nearby star-forming regions. The small disk sizes may be a consequence either of external photoevaporation or the intermediate age of the region. 
\end{itemize}
Our observations suggest that substructures are common not only in disks in the mildly irradiated nearby star-forming regions, but also in disks exposed to intermediate levels of external UV radiation ($\sim10^2-10^3$ $G_0$). If these substructures trace planet-disk interactions, ice and gas giants may still be forming on Solar System scales in $\sigma$ Orionis, but giant planet formation at significantly larger semi-major axes ($\sim50-100$ au) may be rarer compared to nearby star-forming regions. These observations motivate high-resolution imaging of disks in more extreme UV environments to investigate the universality of disk substructures. 

\section*{}
 We thank our contact scientist Ryan Loomis for his assistance with the project. We also thank the anonymous referee for comments improving the manuscript. This paper makes use of the following ALMA data: ADS/JAO.ALMA\#2016.1.00447.S, ADS/JAO.ALMA\#2022.1.00728.S. ALMA is a partnership of ESO (representing its member states), NSF (USA) and NINS (Japan), together with NRC (Canada), MOST and ASIAA (Taiwan), and KASI (Republic of Korea), in cooperation with the Republic of Chile. The Joint ALMA Observatory is operated by ESO, AUI/NRAO and NAOJ. This material is based upon work supported by the National Science Foundation under Grant No. 2307916. T.B. acknowledges funding from the European Union under the European Unionʼs Horizon Europe Research and Innovation Programme 101124282 (EARLYBIRD) and funding by the Deutsche Forschungsgemeinschaft (DFG, German Research Foundation) under grant 325594231, and Germany's Excellence Strategy - EXC-2094 - 390783311. Views and opinions expressed are, however, those of the authors only and do not necessarily reflect those of the European Union or the European Research Council. Neither the European Union nor the granting authority can be held responsible for them. Support for F.L. was provided by NASA through the NASA Hubble Fellowship grant \#HST-HF2-51512.001-A awarded by the Space Telescope Science Institute, which is operated by the Association of Universities for Research in Astronomy, Incorporated, under NASA contract NAS5-26555S.Z. acknowledges support through the NASA FINESST grant 80NSSC20K1376. Support for S.Z. was provided by NASA through the NASA Hubble Fellowship grant \#HST-HF2-51568 awarded by the Space Telescope Science Institute, which is operated by the Association of Universities for Research in Astronomy, Inc., for NASA, under contract NAS5-26555.

\facilities{ALMA}

\software{\texttt{analysisUtils} \citep{2023zndo...7502160H}, 
\texttt{AstroPy} \citep{2013AA...558A..33A, 2018AJ....156..123A, 2022ApJ...935..167A}, 
\texttt{CASA} \citep{2022PASP..134k4501C}, 
\texttt{cmasher} \citep{cmasher}, 
\texttt{eddy} \citep{2019JOSS....4.1220T},
\texttt{frankenstein} \citep{2020MNRAS.495.3209J},
\texttt{Interative Distance Estimation} \url{https://github.com/ElisaHaas25/Interactive-Distance-Estimation/tree/main},
\texttt{linmix} (\url{https://github.com/jmeyers314/linmix}, ported from \citealt{2007ApJ...665.1489K}),
 \texttt{matplotlib} \citep{Hunter:2007},  
 \texttt{mpol} \citep{2023PASP..135f4503Z, mpol},
 \texttt{pandas} \citep{the_pandas_development_team_2022_7344967},
 \texttt{pyro} \citep{JMLR:v20:18-403}, 
 \texttt{pytorch} \citep{pytorch},
 \texttt{visread} \citep{ian_czekala_2021_4432520},
 \texttt{SciPy} \citep{2020SciPy-NMeth}}

\appendix
\section{Model selection \label{sec:modelselection}}
The Bayesian Information Criterion (BIC) is a metric that is used to select between different models for a given set of observations 
\citep{Schwarz1978}. It is defined as $\text{BIC} = k\ln n - 2 \ln L (\hat\theta)$ where $k$ is the number of free parameters in the model, $n$ is the number of datapoints, and $L(\hat\theta)$ is the likelihood evaluated at the posterior median values. A model is favored if its BIC value is lower than that of another model. Thus, the BIC penalizes more complex models unless the likelihood substantially increases.  

The expression that we use for our radial profiles, given in Equation \ref{eq:radprofile}, has $N+1$ terms, where $N$ can vary. While the value of $N$ we initially chose to model each disk is based on visual inspection of the observed radial profile, for systems where a discrepancy between the model and observations is apparent, we then ran models for higher values of $N$ and calculated the corresponding BIC. For each of the following disks, we report the BICs for the different values of $N$ tested in Table \ref{tab:BIC}. Since it is the relative rather than the absolute BIC value that matters for model selection, we subtract a constant for each disk such that the lowest scoring model has a BIC of 0. 

\begin{deluxetable}{ccc}[h]
\tablecaption {BIC values \label{tab:BIC}}
\tablehead{\colhead{Source}&\colhead{$N$} &\colhead{BIC}}
\startdata
SO 662 & 0 & 69 \\
       & 1 & 0 \\
\hline 
SO 844 & 2 & 26 \\
       & 3 & 0 \\
\hline
SO 984 & 0 & 371\\
       & 1 & 0\\
\hline
SO 1036 & 0& 281 \\
        & 1 & 0 \\
\hline
SO 1152 & 2& 0 \\
        & 3& 143 \\
        & 4& 100 \\
\hline
SO 1153 & 2 & 90\\
        & 3 & 0\\
\hline
SO 1274 & 5 & 278\\
        & 6 & 98\\
        & 7 & 0
\enddata 
\end{deluxetable}

\begin{figure}[h]
\begin{center}
\includegraphics{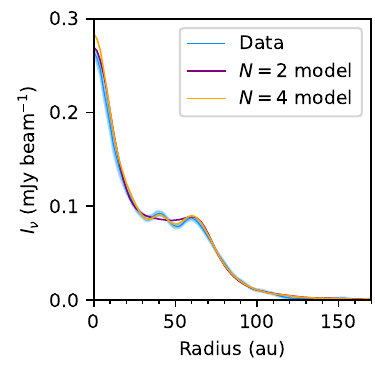}
\end{center}
\caption{A comparison of the $N=2$ and $N=4$ model radial profiles (calculated from the CLEAN images made from the model visibilities) to the observed profile of the SO 1152 disk.  \label{fig:SO1152SVI}}
\end{figure}

For most disks, the final value of $N$ we selected was determined by whichever yields the lowest BIC. However, for SO 1152, we selected the model with $N=4$ over the model with $N=2$ because the latter does not reproduce the rings (Figure \ref{fig:SO1152SVI}). The $N=4$ model overpredicts the peak intensity compared to the $N=2$ model, but since we are primarily interested in characterizing the gaps and rings, we used the $N=4$ model for analysis. 

\section{Parametric model values 
\label{sec:modelparameters}}
Table \ref{tab:modelparameters} lists the medians of the marginalized posteriors and the $68\%$ confidence intervals for the parametric radial intensity models from Section \ref{sec:analysis}. 
\startlongtable
\begin{deluxetable}{c|c|c}
\tablecaption {Inferred parameters for radial intensity profiles \label{tab:modelparameters}}
\tablehead{\colhead{Source}&\colhead{Parameter} &\colhead{Value}}
\startdata
SO 662 & $\Delta x$ (mas) & $1.3\pm0.3$\\
       & $\Delta y$ (mas) & $-0.1\pm0.4$ \\
       & P.A. ($^\circ$) & $109.6\substack{+1.3\\-1.2}$ \\
       & $i$ ($^\circ$) & $53.4\pm1.1$ \\
       & $A_0$ (Jy arcsec$^{-2}$) &$1.6\pm0.1$\\
       & $\sigma_0$ (au)&$4.0\substack{+0.3\\-0.2}$\\
       & $A_1$ (Jy arcsec$^{-2}$) &$0.50\pm0.02$\\
       & $\sigma_1$ (au)&$3.8\pm0.2$\\
       & $r_1$ (au)& $12.8\pm0.4$ \\
\hline 
SO 844 & $\Delta x$ (mas) & $0.7\pm0.2$\\
       & $\Delta y$ (mas) & $0.6\pm0.3$\\
       & P.A. ($^\circ$) & $109.0\substack{+1.8\\-1.5}$\\
       & $i$ ($^\circ$) & $30.7\substack{+0.7\\-0.6}$\\
       & $A_0$ (Jy arcsec$^{-2}$) &$1.6\pm0.2$\\
       & $\sigma_0$ (au)&$2.29\substack{+0.17\\-0.14}$\\
       & $A_1$ (Jy arcsec$^{-2}$) & $0.44\pm0.01$\\
       & $\sigma_1$ (au)&$2.8\pm0.1$\\
       & $r_1$ (au)& $20.5\pm0.1$\\
       & $A_2$ (Jy arcsec$^{-2}$) &$0.272\substack{+0.008\\-0.009}$\\
       & $\sigma_2$ (au)&$3.0\pm0.1$\\
       & $r_2$ (au)& $36.8\pm0.2$ \\
       & $A_3$ (Jy arcsec$^{-2}$) &$0.016\substack{+0.006\\-0.004}$\\
       & $\sigma_3$ (au)&$3.4\substack{+1.5\\-1.1}$\\
       & $r_3$ (au)& $46\pm2$\\
\hline
SO 897 & $\Delta x$ (mas) & $-6.3\pm0.5$ \\
       & $\Delta y$ (mas) & $-4.4\pm0.4$ \\
       & P.A. ($^\circ$) & $171.8\pm0.6$\\
       & $i$ ($^\circ$) & $62.5\pm0.6$ \\
       & $A_1$ (Jy arcsec$^{-2}$) & $0.84\pm0.02$\\
       & $\sigma_1$ (au)& $3.9\pm0.1$\\
       & $r_1$ (au)& $11.9\pm0.1$ \\
       & $z_0$ (arcsec) & $0.09\pm0.03$ \\
       & $\phi$ & $1.12\substack{+0.06\\-0.07}$\\
\hline
SO 984 & $\Delta x$ (mas) & $-2.4\pm0.3$\\
       & $\Delta y$ (mas) & $-0.4\substack{+0.2\\-0.3}$\\
       & P.A. ($^\circ$) & $75.0\pm0.6$\\
       & $i$ ($^\circ$) & $51.5\pm0.4$ \\
       & $A_0$ (Jy arcsec$^{-2}$) &$0.89\pm0.02$\\
       & $\sigma_0$ (au)&$10.9\pm0.3$\\
       & $A_1$ (Jy arcsec$^{-2}$) &$0.319\pm0.005$\\
       & $\sigma_1$ (au)&$10.4\pm0.2$\\
       & $r_1$ (au)& $29.7\substack{+0.4\\-0.3}$\\
\hline
SO 1036 & $\Delta x$ (mas) & $0.3\pm0.3$\\
       & $\Delta y$ (mas) & $1.1\pm0.4$\\
       & P.A. ($^\circ$) & $16.2\substack{+1.0\\-1.1}$ \\
       & $i$ ($^\circ$) & $47.2\pm0.6$\\
       & $A_0$ (Jy arcsec$^{-2}$) &$1.04\pm0.04$\\
       & $\sigma_0$ (au)&$6.1\substack{+0.2\\-0.3}$\\
       & $A_1$ (Jy arcsec$^{-2}$) &$0.320\pm0.006$\\
       & $\sigma_1$ (au)&$11.3\pm0.3$\\
       & $r_1$ (au)& $20.3\pm0.5$\\
\hline
SO 1152 & $\Delta x$ (mas) & $3.1\pm0.4$\\
       & $\Delta y$ (mas) & $0.3\pm0.4$ \\
       & P.A. ($^\circ$) & $144.9\pm0.3$\\
       & $i$ ($^\circ$) & $65.8\pm0.2$ \\
       & $A_0$ (Jy arcsec$^{-2}$) &$0.78\pm0.05$\\
       & $\sigma_0$ (au)&$6.2\pm0.3$\\
       & $A_1$ (Jy arcsec$^{-2}$) &$0.213\substack{+0.009\\-0.008}$\\
       & $\sigma_1$ (au)&$12.1\pm0.6$\\
       & $r_1$ (au)&$17.8\pm1.0$\\
       & $A_2$ (Jy arcsec$^{-2}$) &$0.156\pm0.009$\\
       & $\sigma_2$ (au)&$3.2\substack{+0.3\\-0.2}$\\
       & $r_2$ (au)&$41.4\pm0.5$ \\
       & $A_3$ (Jy arcsec$^{-2}$) &$0.125\pm0.003$\\
       & $\sigma_3$ (au)&$7.9\pm0.3$\\
       & $r_3$ (au)& $61.6\substack{+0.5\\-0.4}$\\
       & $A_4$ (Jy arcsec$^{-2}$) &$0.063\pm0.002$\\
       & $\sigma_4$ (au)&$19.8\substack{+0.8\\-0.7}$\\
       & $r_4$ (au)& $64.3\pm1.5$\\
\hline
SO 1153 & $\Delta x$ (mas) & $0.3\pm0.1$ \\
       & $\Delta y$ (mas) & $0.3\pm0.1$ \\
       & P.A. ($^\circ$) & $162.0\pm0.5$ \\
       & $i$ ($^\circ$) & $47.4\pm0.3$ \\
       & $A_0$ (Jy arcsec$^{-2}$) & $5.9\pm0.2$\\
       & $\sigma_0$ (au)& $2.46\pm0.06$\\
       & $A_1$ (Jy arcsec$^{-2}$) & $0.41\pm0.01$\\
       & $\sigma_1$ (au)& $5.2\pm0.1$\\
       & $r_1$ (au)&$17.0\pm0.2$\\
       & $A_2$ (Jy arcsec$^{-2}$) & $0.235\pm0.002$\\
       & $\sigma_2$ (au)&$17.1\pm0.2$\\
       & $r_2$ (au)& $31.1\pm0.6$\\
       & $A_3$ (Jy arcsec$^{-2}$) &$0.084\pm0.002$\\
       & $\sigma_3$ (au)&$10.2\pm0.3$\\
       & $r_3$ (au)&  $62.0\substack{+0.5\\-0.6}$\\
\hline
SO 1274 & $\Delta x$ (mas) & $-0.8\pm0.2$\\
       & $\Delta y$ (mas) & $-0.3\pm0.3$\\
       & P.A. ($^\circ$) & $135.6\pm2.7$ \\
       & $i$ ($^\circ$) & $16.0\pm0.7$ \\
       & $A_0$ (Jy arcsec$^{-2}$) &$1.15\pm0.06$\\
       & $\sigma_0$ (au)& $3.7\pm0.2$\\
       & $A_1$ (Jy arcsec$^{-2}$) &$0.308\pm0.008$\\
       & $\sigma_1$ (au)&$7.2\pm0.2$\\
       & $r_1$ (au)&$12.2\pm0.4$\\
       & $A_2$ (Jy arcsec$^{-2}$) &$0.228\pm0.009$\\
       & $\sigma_2$ (au)&$2.4\pm0.1$\\
       & $r_2$ (au)& $32.89\pm0.15$\\
       & $A_3$ (Jy arcsec$^{-2}$) &$0.134\substack{+0.008\\-0.007}$\\
       & $\sigma_3$ (au)&$2.1\pm0.1$\\
       & $r_3$ (au)&$49.7\pm0.2$ \\
       & $A_4$ (Jy arcsec$^{-2}$) & $0.103\pm0.006$\\
       & $\sigma_4$ (au)&$2.6\pm0.2$\\
       & $r_4$ (au)&$65.6\pm0.2$ \\
       & $A_5$ (Jy arcsec$^{-2}$) & $0.066\pm0.004$\\
       & $\sigma_5$ (au)&$3.1\pm0.2$\\
       & $r_5$ (au)&$88.1\pm0.3$ \\
       & $A_6$ (Jy arcsec$^{-2}$) &$0.081\pm0.005$\\
       & $\sigma_6$ (au)&$2.3\pm0.2$\\
       & $r_6$ (au)& $119.3\pm0.3$\\
       & $A_7$ (Jy arcsec$^{-2}$) & $0.0229\pm0.0009$\\
       & $\sigma_7$ (au)&$15.0\pm0.5$\\
       & $r_7$ (au)& $120.9\pm0.8$\\
\enddata 
\end{deluxetable}

\section{Identifying shoulders \label{sec:shoulder}}
\begin{figure*}[h]
\begin{center}
\includegraphics{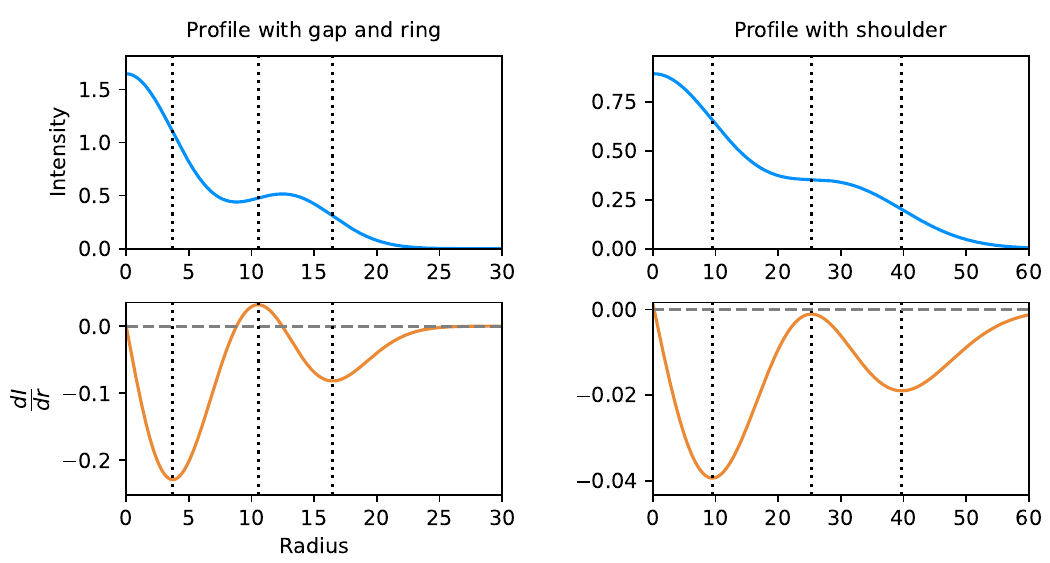}
\end{center}
\caption{\textit{Left}: Intensity and $\frac{dI}{dr}$ profiles for a disk with a gap and a ring. The locations of inflection points in the radial intensity profile are marked with dotted black lines. \textit{Right}: Similar to the left column, but for a disk with a shoulder. \label{fig:shoulder}}
\end{figure*}

In this work, we define a gap as a local minimum and a ring as a local maximum in the model radial intensity profile. Inward of the gap and outward of the ring, there are inflection points marking transitions in the profile from downward to upward concavity. These can be identified as local minima in the $\frac{dI}{dr}$ profile. In between these two inflection points, there is an inflection point separating the gap and ring. This inflection point marks the transition from upward to downward concavity, and can be identified as a local maximum in the $\frac{dI}{dr}$ profile. This behavior is illustrated in the first column of Figure \ref{fig:shoulder}. Along similar lines, we define a shoulder to be present if there is a series of three consecutive inflection points without a gap or ring occurring between them (see the second column of Figure \ref{fig:shoulder}). In this case, the $\frac{dI}{dr}$ profile around the shoulder has a similar shape as the $\frac{dI}{dr}$ profile near a gap-ring pair, but the former never crosses zero before the first and last inflection point. Our definition of a shoulder is similar to that used by \citet{2024PASJ...76..437Y} in their analysis of disks in Taurus, but they also introduce an additional category, ``disk skirt,'' which we do not use for the sake of simplicity. 
\section{Comparison of parametric model profiles to non-parametric model profiles from \texttt{Frank} \label{sec:frank}}

Using the P.A., inclination, and phase center offset derived from the parametric models in Section \ref{sec:analysis}, we generated radial intensity profiles through non-parametric visibility modeling with \texttt{frank} \citep{2020MNRAS.495.3209J}. The peak S/N in the CLEAN images of our targets range from $\sim20-80$, whereas \texttt{frank} has commonly been used to model datasets with peak S/N an order of magnitude higher \citep[e.g.,][]{2021MNRAS.501.2934C, 2022MNRAS.509.2780J, 2022MNRAS.514.6053J}. Consequently, our model intensity profiles are more sensitive to the choice of hyperparameters $\alpha$ (which determines the maximum baseline that will be included in the fit based on its S/N) and $w$ (which determines how much the power spectrum will be smoothed). The recommended values in the code documentation range from 1.05 to 1.3 for $\alpha$ and $10^{-4}$ to $10^{-1}$ for $w$, with larger values being more conservative. We thus ran models with two pairs of values, ($\alpha = 1.05$, $w=10^{-4}$) and ($\alpha=1.3$, $w=0.1$). The fits were performed in logarithmic brightness space because we found that fitting in linear space tended either to produce stronger oscillatory artifacts or negative intensities in some of the deep gaps.

\begin{figure*}[h]
\begin{center}
\includegraphics{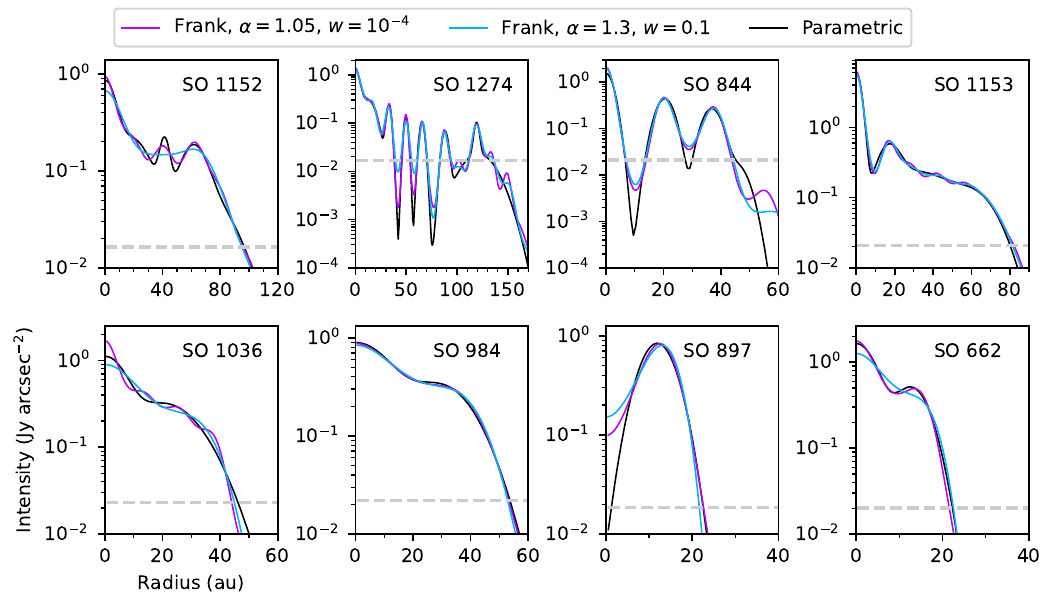}
\end{center}
\caption{A comparison of the model intensity profiles derived from the parametric method (from Section \ref{sec:analysis}) and with \texttt{frank} using two different sets of hyperparameters. The dashed gray lines show the rms of the CLEAN images of the observations.  \label{fig:frankcomparison}}
\end{figure*}

A comparison of the \texttt{frank} models to the parametric models is shown in Figure \ref{fig:frankcomparison}. By and large, the parametric models and \texttt{frank} models recover substructures at the same radial locations. For the deep gaps (i.e., in SO 1274 and SO 897), the parametric models tend to yield deeper gaps compared to the \texttt{frank} models, and the $\texttt{frank}$ model depths are sensitive to the hyperparameter choices. In the SO 1274, SO 1153, and SO 1036 profiles using $\alpha = 1.05$ and $w=10^{-4}$, small-scale oscillations are visible, suggesting that the data are being overfit. 

\begin{figure*}
\begin{center}
\includegraphics{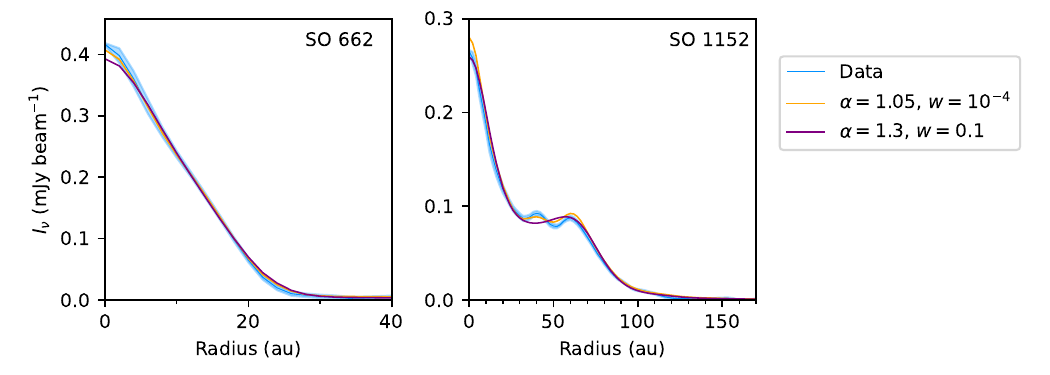}
\end{center}
\caption{Left: A comparison of the \texttt{frank} model radial profiles (calculated from the CLEAN images made from the model visibilities) to the observed profile of the SO 662 disk. Right: Same, but for the SO 1152 disk. \label{fig:SO662frank}}
\end{figure*}

Whereas the ($\alpha = 1.05$, $w=10^{-4}$) model for SO 662 has a gap (similar to the parametric model), the ($\alpha = 1.3$, $w=0.1$) model only has a shoulder. We imaged the model visibilities generated by \texttt{frank} and compared the radial profiles extracted from the model images to the observed radial profile (Figure \ref{fig:SO662frank}). The ($\alpha = 1.05$, $w=10^{-4}$) model better reproduces the radial profile. Thus, both the parametric and non-parametric modeling approaches suggest that a gap is present in the SO 662 disk. 

Neither set of \texttt{frank} models reproduces the gaps and rings in the SO 1152 disk well (Figure \ref{fig:SO662frank}). The previous section noted the difficulty in reproducing the SO 1152 radial profile with the parametric approach as well. This may be a consequence of the vertical structure influencing the observed emission due to the disk's high inclination ($\sim66^\circ$), but deeper observations will be needed to clarify the disk's emission structure.

\end{CJK*}
\end{document}